\documentclass[reprint,prb,amsmath, amssymb, superscriptaddress]{revtex4-1}
\usepackage{times}
\usepackage{verbatim}
\usepackage{graphicx}
\usepackage{graphics}
\usepackage{amsmath}
\usepackage{amssymb}
\usepackage{multirow}
\usepackage{color}
\usepackage{threeparttable}

\begin{document}

\title{Lattice Dynamics of Ultrathin FeSe Films on SrTiO$_3$}

\author{Shuyuan Zhang}
\author{Jiaqi Guan}
\affiliation{Beijing National Laboratory for Condensed Matter Physics and Institute of Physics, Chinese Academy of Sciences, Beijing 100190, China}
\affiliation{School of Physical Sciences, University of Chinese Academy of Sciences, Beijing 100049, China}
\author{Yan Wang}
\affiliation{Department of Physics and Astronomy, University of Tennessee, Knoxville, Tennessee 37996, USA}
\author{Tom Berlijn}
\affiliation{Center for Nanophase Materials Sciences, Oak Ridge
National Laboratory, Oak Ridge, Tennessee 37831, USA}
\affiliation{Computational Sciences and Engineering Division, Oak
Ridge National Laboratory, Oak Ridge,Tennessee 37831, USA}
\author{Steve Johnston}
\affiliation{Department of Physics and Astronomy, University of Tennessee, Knoxville, Tennessee 37996, USA}
\author{Xun Jia}
\author{Bing Liu}
\author{Qing Zhu}
\author{Qichang An}
\author{Siwei Xue}
\affiliation{Beijing National Laboratory for Condensed Matter Physics and Institute of Physics, Chinese Academy of Sciences, Beijing 100190, China}
\affiliation{School of Physical Sciences, University of Chinese Academy of Sciences, Beijing 100049, China}
\author{Yanwei Cao}
\author{Fang Yang}
\author{Weihua Wang}
\affiliation{Beijing National Laboratory for Condensed Matter Physics and Institute of Physics, Chinese Academy of Sciences, Beijing 100190, China}
\author{Jiandi Zhang}
\author{E. W. Plummer}
\affiliation{Department of Physics and Astronomy, Louisiana State University, Baton Rouge, Louisiana 70808, USA}
\author{Xuetao Zhu}
\email[]{xtzhu@iphy.ac.cn}
\affiliation{Beijing National Laboratory for Condensed Matter Physics and Institute of Physics, Chinese Academy of Sciences, Beijing 100190, China}
\affiliation{School of Physical Sciences, University of Chinese Academy of Sciences, Beijing 100049, China}
\author{Jiandong Guo}
\email[]{jdguo@iphy.ac.cn}
\affiliation{Beijing National Laboratory for Condensed Matter Physics and Institute of Physics, Chinese Academy of Sciences, Beijing 100190, China}
\affiliation{School of Physical Sciences, University of Chinese Academy of Sciences, Beijing 100049, China}
\affiliation{Collaborative Innovation Center of Quantum Matter, Beijing 100871, China}
\date{\today}

\begin{abstract}
Charge transfer and electron-phonon coupling (EPC) are proposed to
be two important constituents associated with enhanced
superconductivity in the single unit cell FeSe films on oxide
surfaces. Using high-resolution electron energy loss spectroscopy
combined with first-principles calculations, we have explored the
lattice dynamics of ultrathin FeSe films grown on SrTiO$_3$. We show
that, despite the significant effect from the substrate on the
electronic structure and superconductivity of the system, the FeSe
phonons in the films are unaffected. The energy dispersion and
linewidth associated with the Fe- and Se-derived vibrational modes
are thickness- and temperature-independent. Theoretical calculations
indicate the crucial role of antiferromagnetic correlation in FeSe
to reproduce the experimental phonon dispersion. Importantly, the
only detectable change due to the growth of FeSe films is the
broadening of the Fuchs-Kliewer (F-K) phonons associated with the
lattice vibrations of SrTiO$_3$(001) substrate. If EPC plays any
role in the enhancement of film superconductivity, it must be the
interfacial coupling between the electrons in FeSe film and the F-K
phonons from substrate rather than the phonons of FeSe.

\end{abstract}
\maketitle

\section{Introduction}

The superconducting transition temperature (T$_C$) for monolayer
FeSe films, with the thickness of one unit cell (uc), on
SrTiO$_3$(001) substrate
(1uc-FeSe/STO)\cite{WangQingyan2012-1ucFeSe,ZhangWenhao2014-CPL-1ucFeSe}
is significantly enhanced to $\sim 60-70$ K
\cite{ZhangWenhao2014-PhysRevB-1ucFeSe,liuDefa2012FeSe,LeeJJ2014FeSereplica,TanShiyong2013FeSe}
(even probably up to 109 K \cite{GeJianfeng2015FeSe}) compared to
bulk FeSe (8 K)\cite{hsu2008FeSe}. This discovery serves as a
prototypical example of interfacial T$_C$ enhancement, which has
drawn the attention of the
community\cite{LiDunghai2015CPB,bozovic2014NP,Hoffman2017monolayerFeSeReview,wang2017FeSeReview,Xue2016FeSeReview}.
Although the mechanism of the interfacial T$_C$  enhancement is not
fully understood, electron doping from STO substrates due to the
oxygen vacancies\cite{TanShiyong2013FeSe} or band
bending\cite{Chen2017TEM} at interface is widely believed to be an
indispensable ingredient\cite{HeShaolong2013FeSe,HeJunfeng2014FeSe}.
On the other hand, in the systems such as the intercalated
(Li,Fe)OHFeSe\cite{Chenxianhui2015LiFeOHFeSe,ZhaoLin2015LiFeOHFeSe},
and thick FeSe films/flakes with alkali metal
adatoms\cite{Seo2015KdopedFeSe,CHPwen2016NC,Miyata2015KdopedFeSe,SongCanli2016KdopedFeSe}
or ionic liquid
gating\cite{Shiogai2015gateFeSe,Lei2015gateFeSe,hanzawa2015gateFeSe},
where the electron density of the FeSe layer can reach a value as
high as that in 1uc-FeSe/STO, the T$_C$ is enhanced only up to $\sim
40$ K. Thus, electron doping is not the only contributor to the
increased T$_C$. There must be other interfacial effects involved to
give rise to the extra $\sim 20$ K enhancement in 1uc-FeSe/STO.

From the structural point of view, the tensile strain induced by
lattice
mismatch\cite{WangQingyan2012-1ucFeSe,TanShiyong2013FeSe,PengRui2014PRLFeSe}
leads to the formation of strain
strips\cite{WangQingyan2012-1ucFeSe}. The specific interfacial
structure with a double-TiO$_x$ termination has been observed by
scanning transmission electron
microscopy\cite{LiFangsen2015FeSe-structure,ZouPRB2016TiO2doublelayer},
and thus the vibration of Ti-O bond might be crucial. Similar $T_C$
enhancement behavior has also been discovered in 1uc-FeSe grown on
various oxide substrates with Ti-O bonds such as
BaTiO$_3$(001)\cite{PengRui2014NCFeSe},
SrTiO$_3$(110)\cite{Zhang2015FeSeSTO110,ZhouGuanyu2015FeSeSTO110},
anatase TiO$_2$(001)\cite{DingHao2016FeSeTiO2}, and rutile
TiO$_2$(001)\cite{Rebec2017prlTiO2}, all with different lattice
constants and crystal orientations. These results suggest that it is
essential to understand the interfacial lattice dynamics.

From the dynamical point of view, phonons have been studied to elucidate the possible contribution to the interfacial $T_C$ enhancement. Some studies have proposed that Fe- and Se-derived phonons (FeSe phonons) could participate in the interfacial $T_C$ enhancement, which is supported by electron scattering with FeSe phonons in inelastic electron tunneling spectroscopy measurements\cite{TangChenjia2016FeSe}. Electron-phonon coupling (EPC) of FeSe phonons is also addressed by ab initio calculations\cite{SinisaCoh2015FeSeECP,WangYan2016,DYXing2014FeSeSTOEPC}. On the other hand, several investigations have shown that phonons from STO substrate can strongly interact with the electrons in the FeSe film as evidenced by the observations of FeSe band replica\cite{LeeJJ2014FeSereplica,Johnstone2016PRBForwardscattering,LiDunghai2015CPB} in angle resolved photoemission spectroscopy (ARPES) measurements and the penetration of the STO Fuchs-Kliewer (F-K) phonons into FeSe film in high-resolution electron energy loss spectroscopy (HREELS) measurements\cite{ShuyuanZhang2016}. However, none of the above studies provide the lattice dynamical information of FeSe films, such as the phonon energy and linewidth as a function of wavevector, which can be used to understand the EPC related to superconductivity enhancement.

If phonons are involved in the interfacial enhancement of T$_C$,
several essential questions arise: Which phonons provide the major
contribution, the FeSe phonons in the thin film or the STO phonons
in the substrate? As the dipole field generated by the STO phonons
can penetrate into the thin FeSe film\cite{ShuyuanZhang2016}, are the
FeSe phonons affected by the STO lattice and its dynamics? In this
paper, we address these questions by measuring the lattice dynamics
using HREELS. We demonstrate that although the electronic structure
and superconducting behavior vary in films with different
thicknesses, the lattice dynamics (phonon spectra and Debye
temperature) of Fe- and Se-derived phonons remain unchanged (from
1uc to 10uc thickness). Thus FeSe phonons are not the essential component in
the enhanced interfacial superconductivity. Additionally, the
first-principles calculation shows that the antiferromagnetic (AFM)
correlation in FeSe is indispensable to quantitatively reproduce the
experimental phonon dispersions in the ultrathin FeSe films,
suggesting that the magnetic correlation or spin fluctuation is
critical not only in FeSe
bulk\cite{wang2015SpinFluctuation,wang2016SpinFluctuationNC} but
also in 1uc-FeSe films. In contrast, surface F-K phonon modes of the
STO(001) substrate are strongly temperature-dependent, and clearly
broaden after the growth of FeSe films. These results indicate that
the penetrating substrate F-K phonon field, interacting with
electrons therein, provides additional glue for the existing
electron pairing of FeSe.

\section{Experiments and Methods}

Lattice dynamics including phonon spectra\cite{Shukla2003MgB2LinewidthPRL,Gnezdilov2013FeSePhonon,NeutronPRL2009CaFeAs,FeSeDebye2010prb} and Debye temperature\cite{FeSeDebye2010prb,LaOFFeAsCPL2008specificheat,FeSeSpecificheat,Lawless1977Debye,ElasticConstantsSTO}, can be obtained using techniques such as inelastic x-ray scattering\cite{Shukla2003MgB2LinewidthPRL}, Raman scattering\cite{Gnezdilov2013FeSePhonon}, inelastic neutron scattering\cite{NeutronPRL2009CaFeAs}, nuclear inelastic scattering\cite{FeSeDebye2010prb},specific heat\cite{LaOFFeAsCPL2008specificheat,FeSeSpecificheat}, elastic constant\cite{Lawless1977Debye,ElasticConstantsSTO}, \textit{etc}. These techniques, however, only measure the bulk properties of materials. Substrate effects on an individual phonon branch of ultrathin films, such as FeSe, can only be measured utilizing surface-sensitive techniques such as HREELS or inelastic helium atom scattering\cite{benedek2013heliumBook}. Due to the limitation of the polycrystal nature of FeSe bulk in pervious studies\cite{FeSeDebye2010prb,Phelan2009FeSePhononDOS}, the phonon dispersions of FeSe have not been obtained. Here, the combination of a momentum resolved surface-sensitive technique (HREELS) and the growth of high-quality single crystalline FeSe films allows for the first observation of the phonon dispersions.

\subsection{Preparation of Oxide Substrates}

In this study, HREELS measurements were performed on various
different samples, including single crystalline oxides, oxide films,
and ultrathin single crystalline FeSe films grown on single crystalline oxides. The
samples and the preparation methods are summarized in
Table.\ref{tb2}. The Nb-doped (0.5\%) SrTiO$_3$ substrates are
annealed at 600$^{\circ}$C for 12 h and then at 950$^{\circ}$C
for 1 h in ultra-high vacuum (UHV) condition, which is labeled as
"treated STO" in the paper and is used as the substrate of FeSe
films. The treated STO samples are covered by thick amorphous
selenium layer at room temperature, and annealed at 600 $^{\circ}$C
to remove the Se capping layer before HREELS measurements. To
compare with treated STO, another Nb-doped (0.5\%) SrTiO$_3$
substrate is etched by HF and annealed at 600 $^{\circ}$C for 12
h before EELS measurement, which is labeled as "clean STO" in
this paper. 40uc SrTiO$_3$ films without Nb doping are grown by
pulsed laser deposition (PLD) at 600$^{\circ}$C in UHV condition and
labeled as "clean STO (w/o Nb)". 40uc BaTiO$_3$ films are grown by
PLD at 670$^{\circ}$C in UHV condition. Both clean STO (w/o Nb) and
BaTiO$_3$ films are annealed at 600 $^{\circ}$C for 12 h before
EELS measurement. TiO$_2$(110) substrates are annealed at
600$^{\circ}$C for 12 h and then at 900$^{\circ}$C for 1 min
in UHV condition. The SrTiO$_3$(110) substrates with two different
surface reconstructions, 2$\times$8 and 4$\times$1 superlattices
respect to the lattice of STO(110) surface, are prepared using the
method in Ref.\onlinecite{CaoYanweiSTO110}.

\begin{table*}
 \centering
\caption{Samples used in the HREELS study.
\label{tb2}}\vspace{0.02in}
 \begin{tabular}{c|c|c|p{5cm}}\hline\hline
     Labels         & Substrate                     & Films                      & Substrate Preparation    \\ \hline\hline
 treated STO        & 0.5\% Nb-doped SrTiO$_3$(001) & -                          & 950$^{\circ}$C annealing and Se treatment \\
 clean STO          & 0.5\% Nb-doped SrTiO$_3$(001) & -                          & HF etching and 600$^{\circ}$C annealing \\
 clean STO (w/o Nb) & 0.5\% Nb-doped SrTiO$_3$(001) & 40uc SrTiO$_3$(001) film (w/o Nb) & HF etching and 600$^{\circ}$C annealing \\
 1uc-FeSe/STO       & 0.5\% Nb-doped SrTiO$_3$(001) & 1uc FeSe films             & 950$^{\circ}$C annealing and Se treatment \\
 3uc-FeSe/STO       & 0.5\% Nb-doped SrTiO$_3$(001) & 3uc FeSe films             & 950$^{\circ}$C annealing and Se treatment \\
 10uc-FeSe/STO      & 0.5\% Nb-doped SrTiO$_3$(001) & 10uc FeSe films            & 950$^{\circ}$C annealing and Se treatment \\
 TiO$_2$(110)(w/o Nb) & rutile TiO$_2$(110)           & -                          & 900$^{\circ}$C annealing \\
 BaTiO$_3$ (w/o Nb) & 0.5\% Nb-doped SrTiO$_3$(001) & 40uc BaTiO$_3$(001) film (w/o Nb) & HF etching and 600$^{\circ}$C annealing\\
 STO(110) 2$\times$8 & 0.5\% Nb-doped SrTiO$_3$(110) & -                 & Ar$^+$ ion sputtering and 1470$^{\circ}$C annealing \\
 STO(110) 4$\times$1 & 0.5\% Nb-doped SrTiO$_3$(110) & -                 & Ar$^+$ ion sputtering and 1470$^{\circ}$C annealing \\
 \hline\hline
 \end{tabular}
 \end{table*}

\subsection{Growth of FeSe films}

\begin{figure*}
\begin{center}
\includegraphics[width=0.6\textwidth]{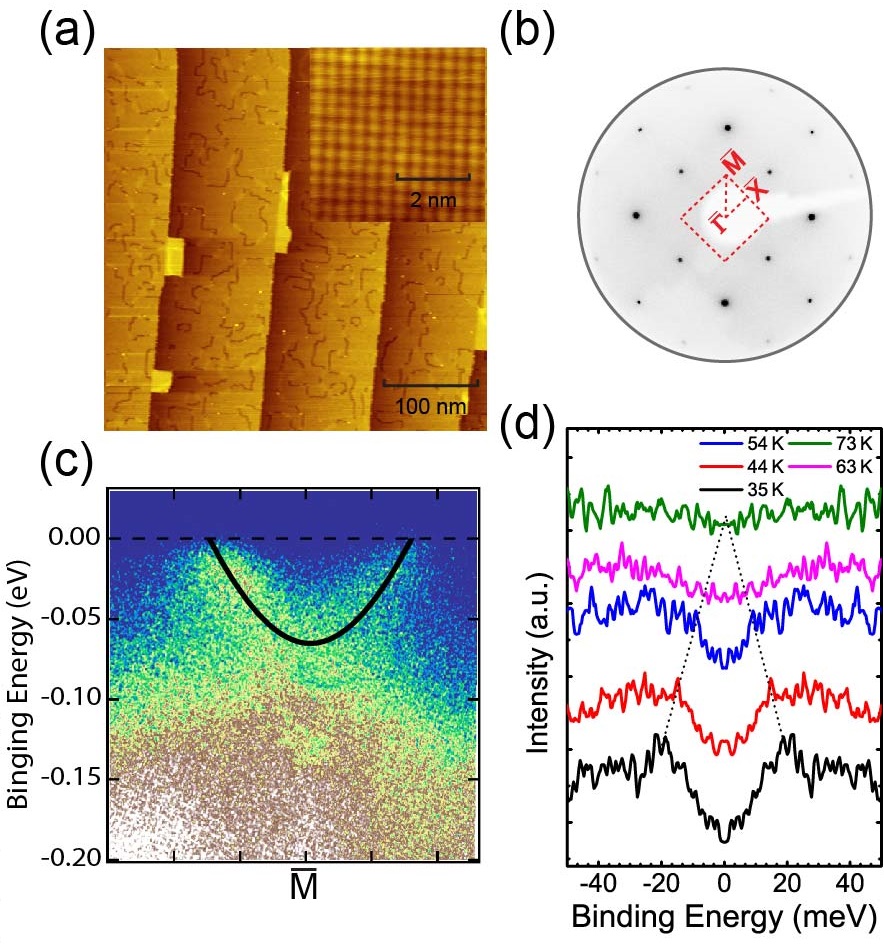}
\caption{\label{fig1} {\small (a) STM topography of 1uc-FeSe/STO
sample ($400\times400$ nm, 5.0 V/50 pA). inset: Atomically resolved
STM topography of 1uc-FeSe/STO sample ($5\times5$ nm, 0.4 V/100 pA).
(b) LEED patterns of 1uc-FeSe/STO sample taken at 35 K with the
primary energy of 80 eV. Red dashed lines represent the first
surface Brillouin zone. (c) ARPES spectrum of 1uc-FeSe/STO, taken at
35 K along the $\overline{\Gamma M}$ direction and centered at
$\overline{M}$. The photon energy is 21.2 eV. The black line
represents the band fitting by a tight binding
model\cite{LeeJJ2014FeSereplica}. (d) Plot of the evolution of the
ARPES symmetrized energy distribution curves (EDCs) near $k_F$ as a
function of temperature, indicating the gap closes between 63 K and
73 K.}}
\end{center}
\end{figure*}

High-quality single crystalline FeSe films were grown by co-depositing high-purity Fe (99.99\%) and Se (99.99+\%) with a flux ratio of $\sim 1:20$ onto the treated STO held at 400 $^\circ$C. The as-prepared samples were post-annealed at 470 $^\circ$C for 5 h in UHV to make the first layer FeSe superconducting. The \emph{in-situ} scanning tunneling microscopy (STM) measurements were performed to confirm the sample quality [Fig. \ref{fig1} (a)]. 1uc, 3uc and 10uc-FeSe/STO samples were capped with thick amorphous Se layer at 300 K and transferred to our two-dimensional (2D) HREELS system\cite{ZhuXuetao2015HREELS}. The Se capping layer was removed by \emph{in-situ} annealing at 450 $^{\circ}$C for 1uc-FeSe/STO and 400 $^{\circ}$C for 3uc and 10uc-FeSe/STO. ARPES measurements were performed for 1uc-FeSe/STO to determine the superconducting gap and the superconducting transition temperature $T_C \sim 65$ K. The Debye temperature measurements were performed using low energy electron diffraction (LEED) on 1uc and 10uc-FeSe/STO samples.

\subsection{ARPES Measurements}

Angle resolved photoemission spectroscopy (ARPES) measurements were performed on 1uc-FeSe/STO samples using a He lamp (21.2 eV) and our 2D-HREELS analyzer along the $\overline{\Gamma M}$ direction, corresponding to the horizontal direction in the LEED pattern in Fig. \ref{fig1}(b). A parabolic electron band can be clearly observed in the spectrum [Fig. \ref{fig1}(c)]. Temperature-dependent EDCs symmetrized around $k_{F}$ reveal the superconducting gap at 35 K is $\sim 20$ meV, which closes between 63 K and 73 K [Fig. \ref{fig1}(d)].

\subsection{Surface Debye Temperature Measurements}

LEED has been used routinely to determine the surface Debye temperature\cite{Tabor1971LEEDDebye,Jones1966Debye,reid1970debye}. According to the Debye-Waller theory\cite{Debye1913,Waller1923}, the coherent peak intensity will decay exponentially with increasing temperature due to the thermal vibrations. The time-averaged scattered intensity affected by thermal vibrations is given by\cite{DebyeWaller1952james}
\begin{align}\label{eq1}
\left\langle I \left( S \right)\right\rangle &=\left\langle I_{\textup{Bragg}}\right\rangle+\left\langle I_{\textup{D}}\right\rangle \\ \nonumber &=\left|f_0\right|^2e^{-2M}\sum\nolimits_i\sum\nolimits_je^{iS\cdot\left(r_i-r_j\right)}+\left\langle I_{\textup{D}}\right\rangle,
\end{align}

with

\begin{equation}\label{eq2}
M=8\pi^2\left\langle u^{2} \right\rangle\frac{sin^2\phi}{\lambda^2},
\end{equation}
and
\begin{equation}\label{eq3}
\left\langle u^{2} \right\rangle\ = \frac{3\hbar^{2}T}{mk_{B}\Theta^2},
\end{equation}

where $f_0$ is the structure factor, $S$ is the diffraction vector $(k-k_0)$, $\left\langle u^{2} \right\rangle$ is the mean square displacement of the atoms$'$ thermal vibrations from their equilibrium position parallel to $S$, $\lambda$ is the wavelength of the scattered radiation, $\phi$ is the Bragg angle, $k_B$ is the Boltzmann constant, $m$ is the atomic mass, and $\Theta$ is the Debye temperature.

The first term $\left\langle I_{\textup{Bragg}}\right\rangle$ in Eq. \ref{eq1} shows that the Bragg intensity is reduced by a prefactor $e^{-2M}$. This prefactor is known as the Debye-Waller factor, containing the effect of the thermal vibrations of the atoms about their equilibrium position. Since the mean square of displacement from equilibrium will increase with increasing temperature [as shown in Eq. \ref{eq2} and Eq. \ref{eq3}, $M$ is proportional to $T$], the thermal vibrations reduces exponentially the intensity of the Bragg peaks\cite{ashcroft1976solid}. Accordingly, the temperature dependence of LEED spot intensity can be used to determine the Debye-Waller factor and calculate the surface Debye temperature. The second term $\left\langle I_{\textup{D}}\right\rangle$ in Eq. \ref{eq1} represents the first order temperature diffuse scattering, which induces a background in LEED pattern and should be subtracted before calculating the Debye temperature.

\subsection{HREELS Measurements}

As a surface-sensitive technique, HREELS is an ideal candidate to study the interfacial lattice dynamics of FeSe/STO systems. Compared with conventional HREELS, our recently developed 2D-HREELS system\cite{ZhuXuetao2015HREELS} can directly map a 2D energy-momentum dispersion over a very large momentum range without mechanically rotating sample, monochromator, or analyzer. Phonon spectra measurements were performed by the 2D-HREELS on 1uc, 3uc and 10uc-FeSe/STO samples.

In HREELS, a monochromic electron beam with energy $E_i$ incident on the sample surface may interact with surface elementary excitations such as phonons, and be scattered with the final energy $E_f$. The energy loss $E_{loss}=E_i-E_f$ represents the energy of surface excitations. In our HREELS measurements, the incident electron beam energies are $E_i$=50 eV and 80 eV for the scattering direction along the $\overline{\Gamma X}$ and $\overline{\Gamma M}$ directions, respectively, and the incident angle($\theta_i$) is 60$^\circ$ with respect to the surface normal. The surface phonon momentum can be determined from the scattering angles by $q=\frac{\sqrt{2m_eE_i}}{\hbar}\left(\sin\theta_i-\sin\theta_f\right)$ when E$_{loss}$ $\ll$ E$_i$, where $m_e$ is the electron mass, $\theta_i$ and $\theta_f$ are the angles of the incident and scattered electrons, respectively. The energy and momentum resolutions of HREELS in this study are $\Delta E \sim$ 3 meV and $\Delta k \sim 0.01 {\AA}^{-1}$, respectively.

For most samples, the HREELS measurements were carried out along both $\overline{\Gamma X}$ and $\overline{\Gamma M}$ directions with the sample temperature ranging from 35 K to 300 K.

\subsection{First Principles Calculation Details}

For the calculations of FeSe phonon dispersions, we use the frozen-phonon method as implemented in the phonopy code \cite{Togo2015} by performing density functional theory (DFT) calculations to extract the interatomic force constant matrix. The DFT calculations are performed using the plane wave projector augmented wave (PAW) method \cite{Bloechl1994} as implemented in the VASP code~\cite{Kresse1996,Kresse1999}. The phonon dispersion calculated with the frozen-phonon method and VASP code (FP/VASP) is
compared with the phonon dispersion calculated with the density functional perturbation theory (DFPT) method as implemented in the
Quantum \textsc{espresso} package\cite{Giannozzi2009} (DFPT/QE), and they agree with each other well with a maximum difference less than $1$~meV. In the DFPT/QE calculations, we use the plane wave ultrasoft pseudopotential method and the GBRV pseudopotential
library\cite{Garrity2014}. In both FP/VASP and DFPT/QE calculations, we take the exchange-correlation functional in the generalized gradient approximation (GGA) of Perdew-Burke-Ernzerhof (PBE) type\cite{Perdew1996}.

In experiment the monolayer FeSe thin film is deposited on the
SrTiO$_3$ substrate, which exerts strong tensile strain on the
monolayer FeSe. To take into account this effect, in the
calculations we set the in-plane lattice parameter $a=3.905$~{\AA}
for the monolayer FeSe thin film, the same as the lattice parameter
for a bulk SrTiO$_3$ crystal. Lattice constants corresponding to the
3 and 10 unit cell FeSe films are extracted from
Ref.\onlinecite{PengRui2014PRLFeSe}. We place a vacuum layer around
$12$~{\AA} in height above the monolayer FeSe before it is repeated
in the $c$ direction. Before phonon dispersion calculations, the
internal atomic coordinates are relaxed until a force smaller than
$1$~meV/{\AA} is found on each atom. We have performed phonon
dispersion calculations for both nonmagnetic (NM) and checkerboard
antiferromagnetic (cAFM) phases and find that the phonon dispersion
of the cAFM phase quantitatively agrees with the EELS experiment
much better than that of NM phase. For the VASP monolayer FeSe
calculations $15\times 15 \times 1$ k-point grids were used to relax
the atoms in the unit cells and $3\times 3 \times 1$ k-grids were
used in $3\times 3 \times 1$ supercells to compute the phonon
dispersions. For the VASP bulk FeSe calculations $15\times 15 \times
10$ k-point grids were used to relax the atoms in the unit cells and
$3\times 3 \times 3$ k-grids were used in $3\times 3 \times 2$
supercells to compute the phonon dispersions. For all VASP
calculations a 500 eV kinetic energy cutoff was employed.

\section{Results}

\subsection{Surface Phonon Spectra}

\begin{figure*}
\begin{center}
\includegraphics[width=0.7\textwidth]{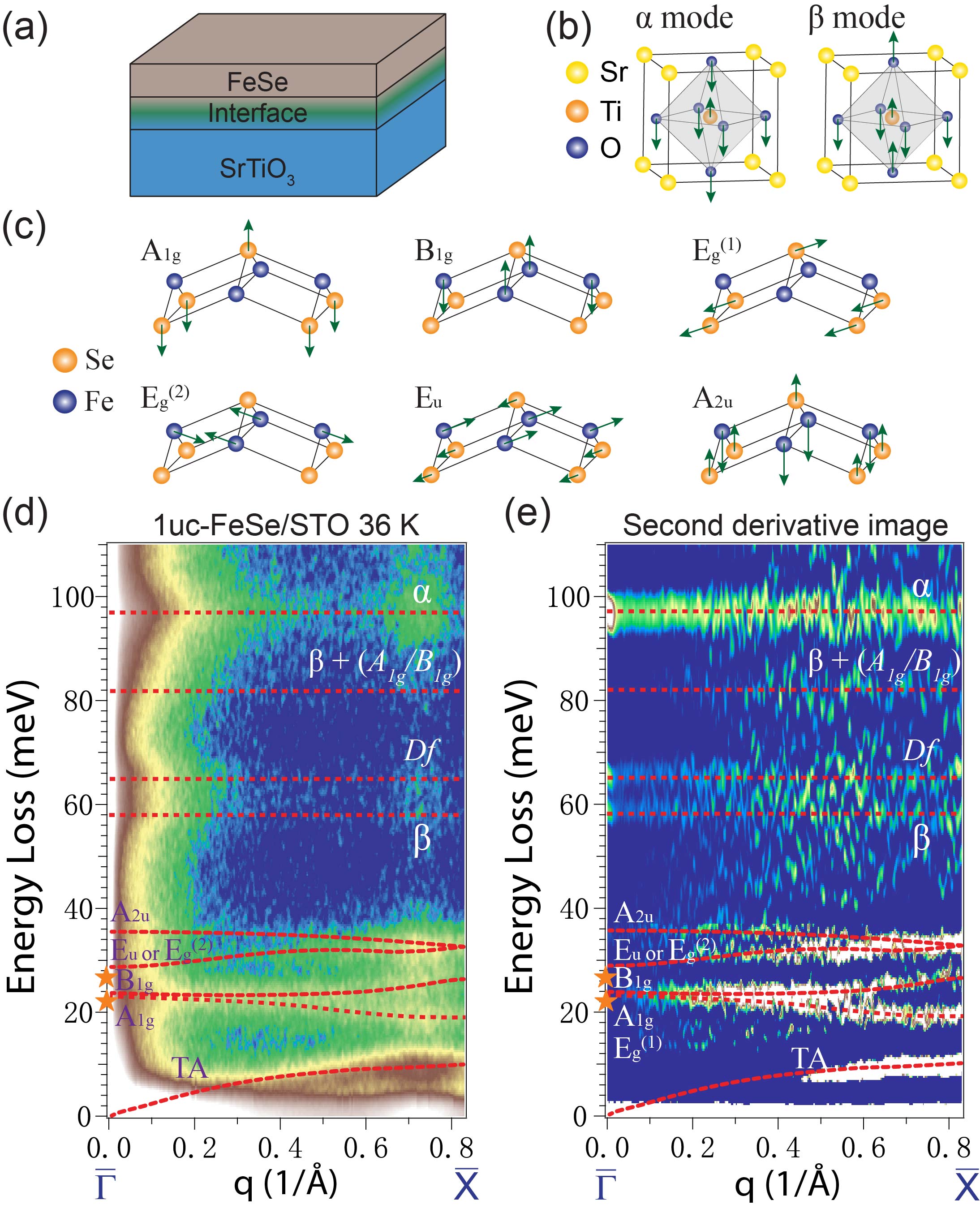}
\caption{\label{fig7} {\small (a) Schematic structure of the FeSe
film on STO substrate. (b) Illustration of ionic vibrations of F-K
phonons in STO. (c) Illustration of ionic vibrations of phonons in
FeSe at $\overline{\Gamma}$ point. (d) Energy-momentum mapping of 2D
HREELS measurements of 1uc-FeSe/STO samples along $\overline{\Gamma
X}$ direction at 35 K, where red solid lines are guides to the eye.
Orange stars label the A$_{1g}$ mode (22.6 meV) and B$_{1g}$ mode
(25.6 meV) measured by Raman scattering at 7
K\cite{Gnezdilov2013FeSePhonon}. (e) Second derivative image of
(d).}}
\end{center}
\end{figure*}

The HREELS results are summarized in Fig. \ref{fig7}.
Figure \ref{fig7}(a) is a schematic drawing of the FeSe layer on
SrTiO$_3$. The FeSe lattice is consisted of Se-Fe-Se triple layers
stacked by van der Waals forces, with the primitive unit cell
containing two Fe atoms and two Se atoms. In each Se-Fe-Se triple
layer, Fe atoms form a square lattice and Se atoms staggered above
and below the Fe plane, as shown in Fig. \ref{fig7}(c). The film with
1uc thickness (~5.5 \AA) corresponds to one Se-Fe-Se triple layer.
Figures \ref{fig7}(b) and \ref{fig7}(c) depict the vibrational modes of some
relevant optical phonons at $q = 0$ of the STO and FeSe,
respectively. To illustrate the phonon dispersions at 1uc-FeSe/STO
surface, in Figs. \ref{fig7}(d) and \ref{fig7}(e), we show a typical 2D-HREELS
energy-momentum mapping for the 1uc-FeSe/STO(001) sample along
$\overline{\Gamma X}$ direction at 36 K.

\begin{figure*}
\begin{center}
\includegraphics[width=0.9\textwidth]{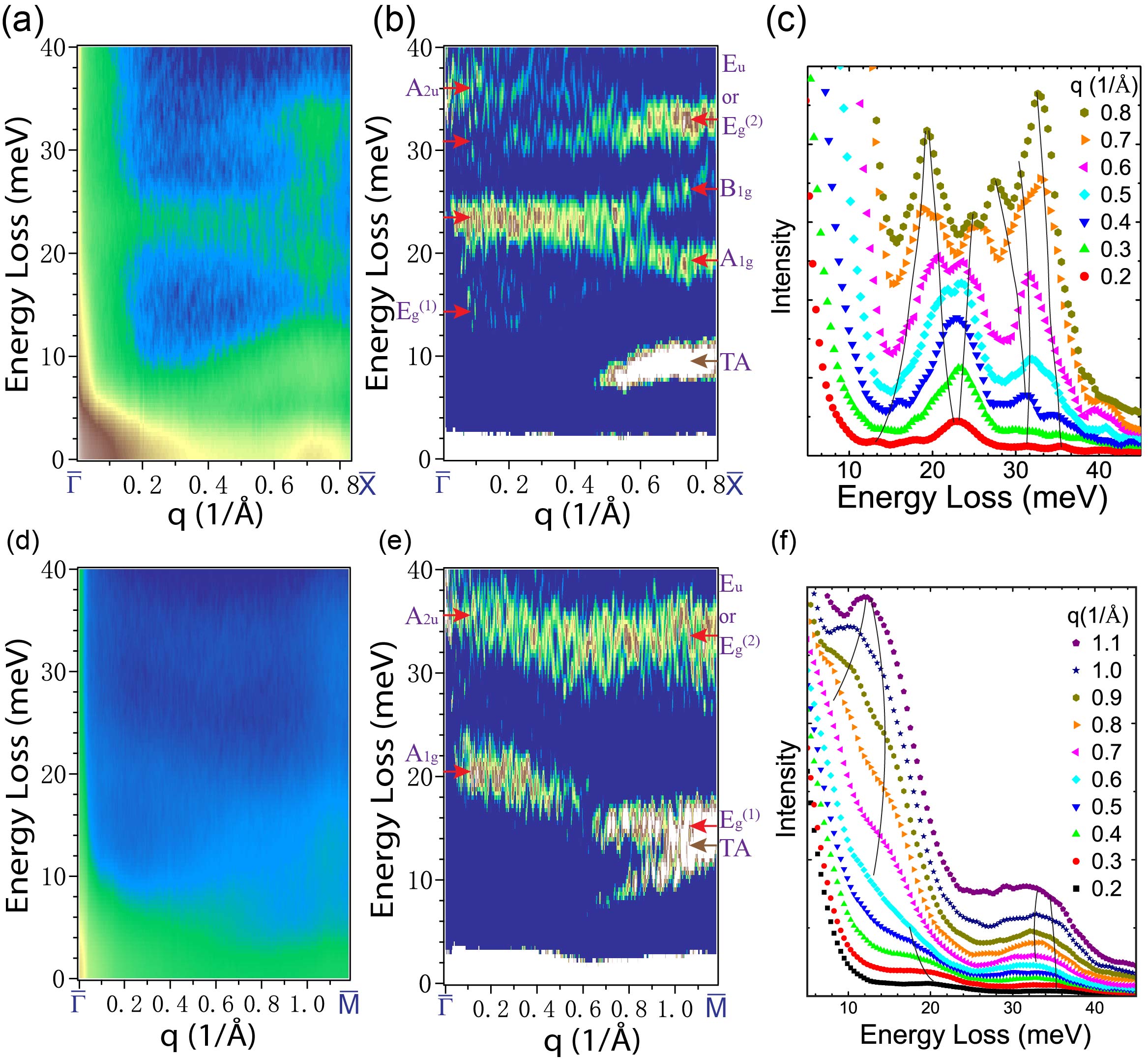}
\caption{\label{fig4} {\small (a) Energy-momentum mapping (0 - 40
meV) of 2D HREELS measurements of 1uc-FeSe/STO samples along
$\overline{\Gamma X}$ direction at 35 K. (b) Second derivative image
of (a). (c) Momentum dependent EDCs of (a), where black lines are
only guides. (d) - (f) Corresponding results along $\overline{\Gamma
M}$ direction.}}
\end{center}
\end{figure*}

In a HREELS measurement, there are two scattering
mechanisms\cite{ibach1982eels}. One is referred to as impact
scattering, where the incident electrons are scattered by impacting
with nuclei of the sample. In this mechanism, the penetration depth
or free electron path of incident electrons (with the energy of 50
eV) is nearly three atomic layers, i.e., around one Se-Fe-Se triple
layer (one unit cell thick in c direction of FeSe crystal). In this
context, most the detected signals of FeSe phonons come from the
topmost Se-Fe-Se layer of FeSe films. In the low energy range 0-40
meV (Fig. \ref{fig4}), five Fe- and Se-derived phonon branches are
clearly observed: one acoustic branch with energy from 0 - 8 meV,
and the other four optical branches with energies ranging from 18 - 40 meV. Another branch around 15 meV [$E_{g}(1)$ mode] has a very weak signal that is barely discernible in the second derivative
image from the data at T=35 K. The energies of two bulk phonon modes
determined from Raman scattering
experiment\cite{Gnezdilov2013FeSePhonon} are also labeled at the
$\overline{\Gamma}$ point in Figs. \ref{fig7}(d) and \ref{fig7}(e), indicating
that the corresponding phonon energies of single layer FeSe films
are similar to the bulk modes.

The other mechanism is called dipole scattering, where the incident
electrons are scattered by dipole fields generated by ionic
vibrations. The ultra-sensitivity to F-K phonons in HREELS
measurements is due to this mechanism. In this case, the incident
electrons may scatter outside the surface nuclei position, depending
on the strength of the dipole field. The STO substrate has two
obvious F-K surface phonons, labeled by $\alpha$ and $\beta$, which
can generate strong electric fields and penetrate through FeSe
films. The buried STO signal can be observed because the electrons
in FeSe layers can not completely screen the electric field from the
F-K phonons\cite{ShuyuanZhang2016}.

\subsection{Thickness Independence of FeSe phonons}

\subsubsection{Surface Phonon Dispersions}

\begin{figure*}
\begin{center}
\includegraphics[width=0.9\textwidth]{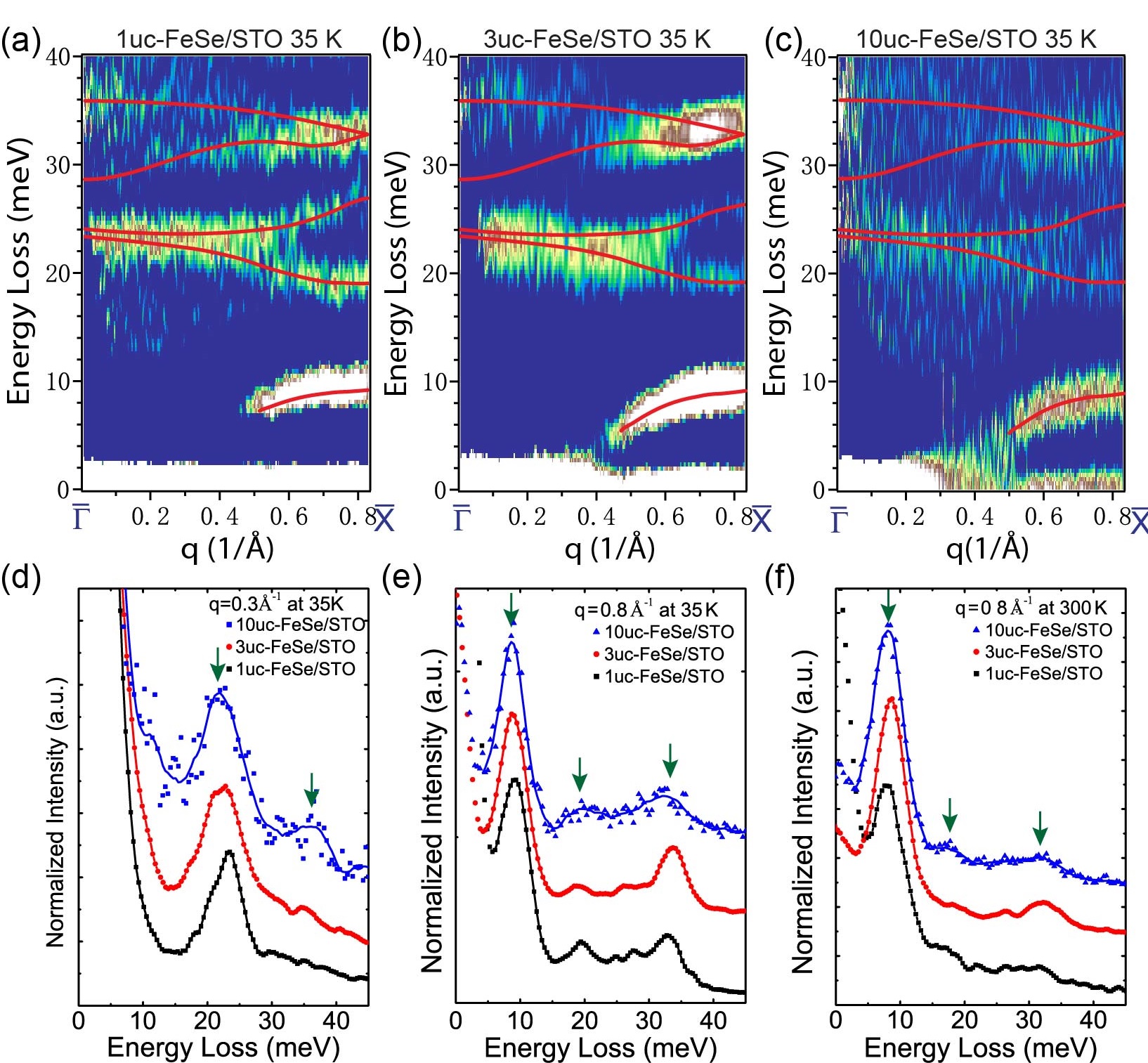}
\caption{\label{fig5} {\small Phonon dispersions of low energy FeSe phonons of (a) 1uc-FeSe/STO, (b) 3uc-FeSe/STO and (c) 10uc-FeSe/STO samples at 35 K, where the colored images are the second derivative energy-momentum mappings of 2D HREELS measurements, and red solid lines are guides to eyes. EDCs from different samples are compared at different momentum points: (d) q $= 0.3$ {\AA}$^{-1}$ at 35 K, (e) q = $0.8$ {\AA}$^{-1}$ at 35K and (f) q $= 0.8$ {\AA}$^{-1}$ at 300K.}}
\end{center}
\end{figure*}

To check the substrate effect on the FeSe phonons as a function of film thickness, phonon spectra of 3uc and 10uc-FeSe/STO are also measured, and compared with 1uc-FeSe/STO in Fig. \ref{fig5}. Phonon energies and dispersions of thick FeSe films show no obvious difference (within the energy resolution) compared to the phonons of 1uc-FeSe films at 35 K. This indicates that charge transfer or other interfacial coupling such as tensile strain does not renormalize the phonon dispersions. As shown by the energy distribution curves (EDCs) at different momentum points in Figs. \ref{fig5}(d) - \ref{fig5}(f), the linewidths of FeSe phonons in different thickness samples are also similar and all close to the instrument resolution. Since the linewidths and dispersion profiles of phonons can reflect the strength of mode-specific EPC\cite{ZhuXuetao2012prlBi2Se3}, the similar linewidth and dispersion for different thickness films provide evidence that the EPC from FeSe phonons is not altered by the existence of substrate or the thickness of the film. Thus, FeSe phonons are not directly related to the interfacial T$_C$ enhancement.

\subsubsection{Surface Debye Temperature}

\begin{figure*}
\begin{center}
\includegraphics[width=0.6\textwidth]{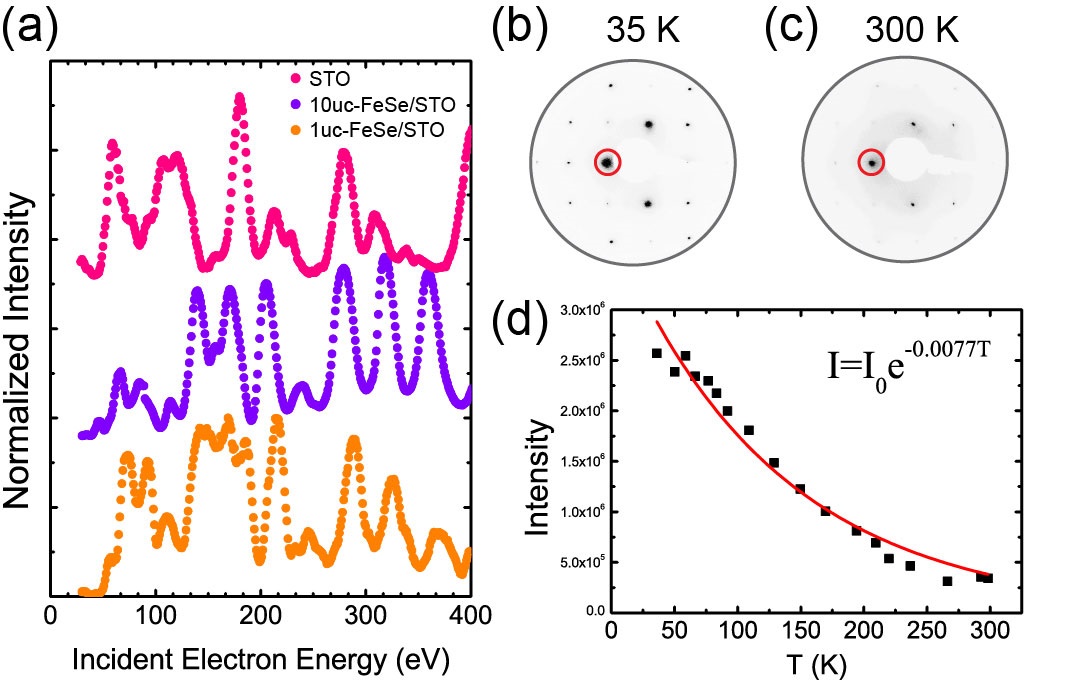}
\caption{\label{fig2} {\small (a) LEED I/V spectra at 35 K of the (00) spot for STO (pink), 1uc-FeSe/STO (purple) and 10uc-FeSe/STO (orange). (b) LEED patterns for 1uc-FeSe/STO at 35 K  and (c) 300 K, for a primary energy of 140 eV. The red circles indicate the (00) spot. (d) The intensity of the (00) spot of 1uc-FeSe/STO as a function of temperature with the primary energy of 220 eV.}}
\end{center}
\end{figure*}

The thickness independence of FeSe phonons is further confirmed by
the measurements of the surface Debye temperature. To check the
lattice contribution in electron pairing, Debye temperature
($\Theta$) is a vital factor in the McMillan
equation\cite{McMillanEquation} and sets the energy scale for $T_C$
in the standard BCS approach in bulk iron
pnictides\cite{LaOFFeAsCPL2008specificheat,Boeri2008ionEPC}. We use
the surface sensitive probe LEED to determine the surface Debye
temperature (see Section II D for details of this technique).

We performed LEED I/V measurements to determine the positions of the Bragg peaks (Bragg positions) for three different samples: STO, 1uc-FeSe/STO and 10uc-FeSe/STO as shown in Fig. \ref{fig2}(a). To acquire the Debye-Waller factor, LEED I/V spectra measurements were performed at different temperatures from 35 - 300 K. Figures \ref{fig2}(b) and \ref{fig2}(c) show the LEED patterns of 1uc-FeSe/STO at 35 K and 300 K respectively. Clearly, the spot intensity at 35 K [Fig. \ref{fig2}(b)] is much stronger than that at 300 K [Fig. \ref{fig2}(c)] because of the reduced thermal vibrations. The spot intensities as a function of temperature at a given voltage (Bragg positions) are plotted and fitted exponentially with $I \left( T \right) \propto e^{-2M}$ to determine the Debye temperature. As an example of the analysis, the LEED (00) spot intensity for 1uc-FeSe/STO at 220 eV is shown in Fig. \ref{fig2}(d), giving $\Theta(220 eV) = 253$ K. In the fitting, we approximate the Bragg angle $\phi$ as 90$^{\circ}$, and $m$ the average atomic mass of FeSe. The average surface Debye temperature measured from different Bragg positions is $\Theta_{1\textup{uc}} \sim 249\pm33$ K for 1uc-FeSe/STO and $\Theta_{10\textup{uc}}\sim 230\pm33$ K for 10uc-FeSe/STO, showing no obvious difference within the statistical error. The surface Debye temperatures for the two different film thicknesses are both located in the range of reported values for bulk FeSe, from 210 K (measured by specific heat\cite{FeSeSpecificheat}) to 285 K (measured by $^{57}$Fe nuclear inelastic scattering\cite{FeSeDebye2010prb}).

\subsection{Changes of Substrate Phonons}

In stark contrast to the unflappable FeSe phonons, the F-K phonons of STO substrate respond to every change in the system. As shown in Fig. \ref{fig10} and Fig. \ref{fig6}, the phonon spectra of the F-K modes for various samples are plotted for comparison. The F-K phonons from the 1uc-FeSe/STO surface show several important features: (1) appearance of new energy loss modes; (2) dramatic temperature dependence; and (3) linewidth broadening compared to the F-K modes of clean STO.

\subsubsection{Appearance of new energy loss modes}

\begin{figure*}
\begin{center}
\includegraphics[width=0.95\textwidth]{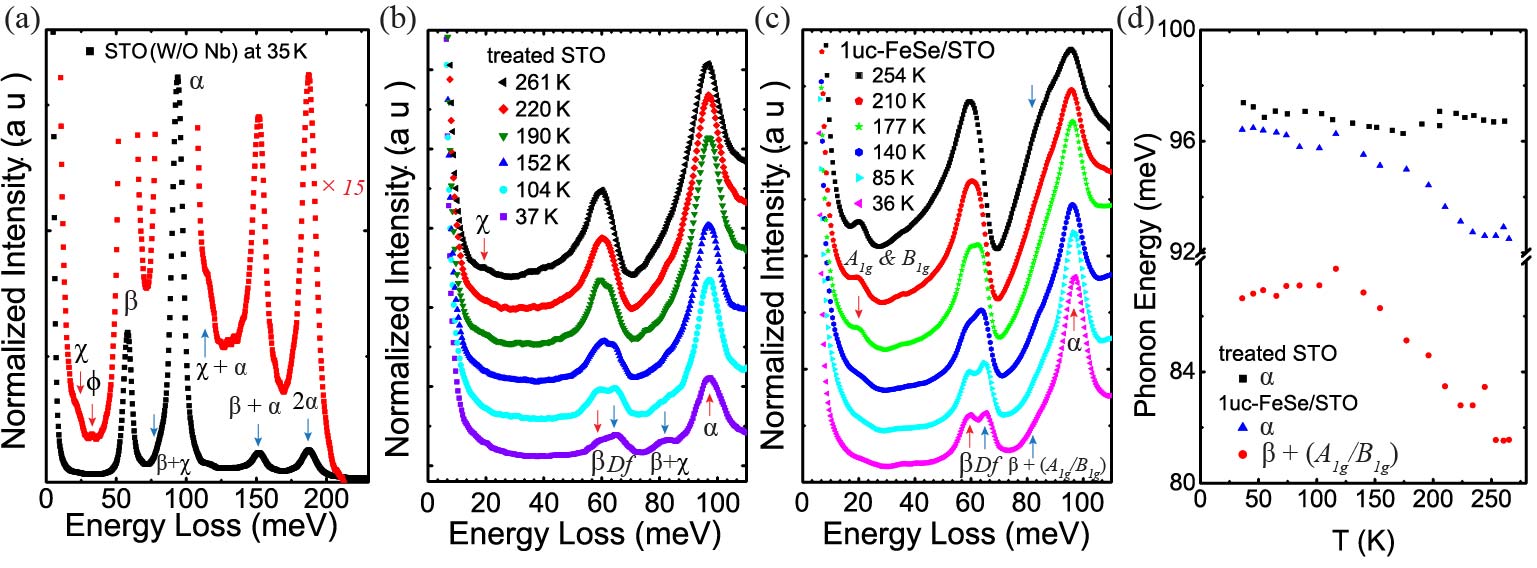}
\caption{\label{fig10} {\small (a) EDCs of phonon spectra at $\overline{\Gamma}$ point of clean STO(001) (w/o Nb) at 35K. Red data are expanded by 15 times from the raw data in black. (b) Temperature-dependent EDCs of treated STO at $\overline{\Gamma}$ point. (c) Temperature-dependent EDCs of 1uc-FeSe/STO at $\overline{\Gamma}$ point. (d) Plot of the energies of several modes as a function of temperature for treated STO and 1uc-FeSe/STO at $\overline{\Gamma}$ point.}}
\end{center}
\end{figure*}

\begin{figure*}
\begin{center}
\includegraphics[width=0.8\textwidth]{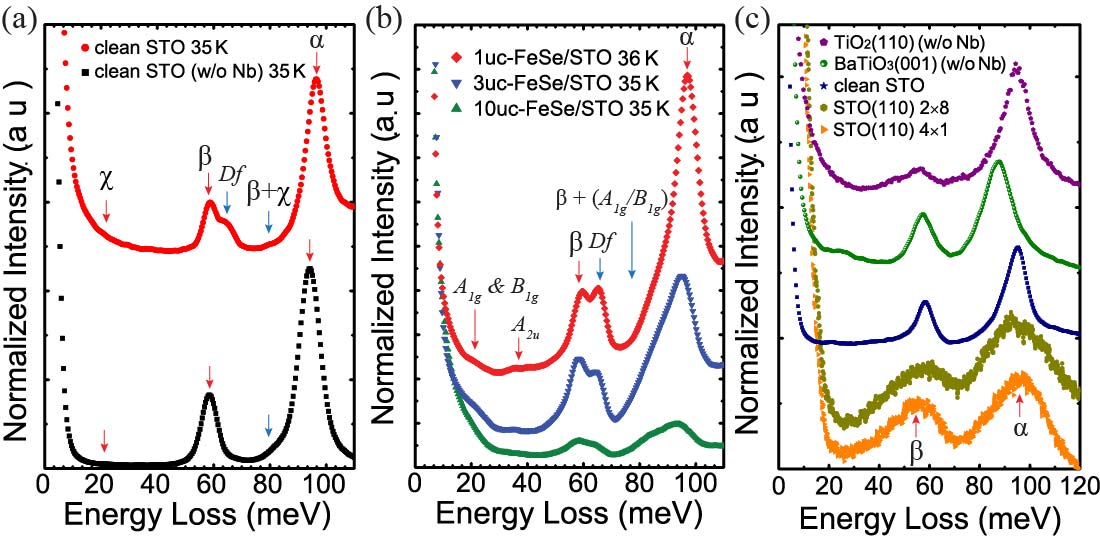}
\caption{\label{fig6} {\small (a) Comparison of EDCs of phonon spectra at $\overline{\Gamma}$ point of clean STO(001) (w/o Nb) and clean STO(001). (b) Comparison of EDCs of phonon spectra at $\overline{\Gamma}$ point of 1uc, 3uc and 10uc-FeSe/STO. (c) Energy loss spectra at $\overline{\Gamma}$ point of several typical oxide surfaces: rutile TiO$_2$(110) (w/o Nb), BaTiO$_3$ (w/o Nb), STO(110) 2$\times$8, STO(110) 4$\times$1, and clean STO, all at room temperature. }}
\end{center}
\end{figure*}

The first feature is the appearance of new energy loss modes, $Df$ and $\beta$ + A$_{1g}$/B$_{1g}$, as shown in Figs. \ref{fig10}(b), \ref{fig10}(c) and Figs. \ref{fig6}(a), \ref{fig6}(b). It turns out that the $Df$ originates from Nb-induced defects in STO substrate and the $\beta$ + A$_{1g}$/B$_{1g}$ mode is an overtone of $\beta$ mode in STO and A$_{1g}$ or B$_{1g}$ mode in FeSe films.

To illuminate the origin of the new energy loss modes, all energy loss modes observed on clean STO (w/o Nb) surface are labeled in Fig. \ref{fig10}(a) and summarized in Table.\ref{tb3}. There are 4 optical phonon modes and 4 overtones. The overtone of $\chi + \beta$ modes can also be observed on treated STO surface as shown in Fig. \ref{fig10}(b). After the growth of FeSe films, A$_1g$ and B$_1g$ modes from Fe- and Se-derived phonons are also involved into the overtone [labeled as $\beta$ + A$_1g$/B$_1g$ in Fig. \ref{fig10}(c)] with the energy similar to $\chi + \beta$ mode. Thus, thicker FeSe films have larger intensity ratio $I( \beta + A_{1g}/B_{1g})/I(\alpha)$, as shown in Fig. \ref{fig6}(b).

For STO with Nb doping, a new energy loss mode with 65.3 meV labeled by $Df$ emerges as shown Fig. \ref{fig10}(b), while this $Df$ mode does not exist in samples without Nb-doping. Thus the $Df$ mode should originate from Nb-induced defects in STO substrate. It is well-known that the superconducting behavior of FeSe films does not depend on Nb-doping of the STO substrate, thus the $Df$ mode is not related to the T$_C$ in 1uc-FeSe/STO samples.

\begin{table*}
 \centering
\caption{The assignments of energy loss modes at $\overline{\Gamma}$ point of clean STO (w/o Nb) surface at 35K.
\label{tb3}}\vspace{0.02in}
 \begin{tabular}{c|c|c|c}\hline\hline
   Phonon Mode & Phonon Energy (meV) & Overtone          & Overtone Energy (meV)   \\ \hline\hline
   $\chi$      & 22.8                & $\chi + \beta$    & 84.2                    \\
   $\phi$      & 32.9                & $\chi + \alpha$   & 115.9                   \\
   $\beta$     & 58.9                & $\beta + \alpha$  & 152.2                   \\
   $\alpha$    & 94.6                & $2\alpha$         & 188.1                   \\
 \hline\hline
 \end{tabular}
 \end{table*}

\subsubsection{Dramatic temperature dependence}

The second feature is the dramatic temperature dependence in phonon energy of F-K phonons as shown in Fig. \ref{fig10}(c). From the comparison between Figs. \ref{fig10}(b) and \ref{fig10}(c), the line profile of the energy loss spectra is different with or without the growth of FeSe films. The difference of temperature-dependent line profile contains two part of contributions: (1) overtone of $\beta$ and A$_1g$/B$_1g$; (2) energy softening caused by anharmonic phonon-phonon interaction.

First, as observed from Figs. \ref{fig10}(b) and \ref{fig10}(c), the intensity of A$_1g$ and B$_1g$ modes in FeSe films increase with increasing temperature, thus the intensity of the overtone $\beta$ + A$_1g$/B$_1g$ becomes stronger at high temperature than that of $\chi + \beta$ on treated STO surface.

Second, on 1uc-FeSe/STO surface [Fig. \ref{fig10}(d)]), the energy of $\alpha$ mode is strongly temperature-dependent, softening from $\sim$ 97 meV at 35 K to $\sim$ 92 meV at 254 K. In contrast, it is almost temperature-independent on all STO substrates without FeSe. This energy shift accompanied with linewidth broadening is due to the anharmonic phonon-phonon interaction, which will lead to the decay of F-K modes into other low energy FeSe phonons. The low energy FeSe phonons provide more possible decay channels for the F-K phonons\cite{ShuyuanZhang2016}. As an example in MgB$_2$\cite{Yildirim2001MgB2}, the anharmonic phonon-phonon interaction can adjust the band structure and then generate strong EPC. Thus, F-K phonons from the STO substrate and interfacial electronic structure might be critical to reveal the interfacial EPC.

\begin{table*}
 \centering
\caption{Energies and line-width of F-K modes at 35K.  $\Gamma_{ep}$ is the linewidth [full width at half maximum (FWHM)] in meV, after deconvoluting the elastic peak width $\Gamma$= 5.6 meV. All the substrates in this table are Nb-doped STO unless stated otherwise.
\label{tb1}}\vspace{0.02in}
\begin{threeparttable}
 \begin{tabular}{c|c|c|c|c|c|c}\hline\hline
            & mode       & clean STO (w/o Nb) & clean STO  & 1uc-FeSe/STO & 3uc-FeSe/STO & 10uc-FeSe/STO  \\ \hline\hline
  Energy (meV)
            & $\alpha$   & 94.1               & 96.5       & 96.9         & 96.4         & 93.4       \\
            & $\beta$    & 59.0               & 58.8       & 59.3         & 58.3         & 58.6       \\  \hline
Linewidth $\Gamma_{ep}$ (meV)
            & $\alpha$   & 6.0                & 7.0   & 7.4-10.2\tnote{1} & 9.7          & 13.7       \\
            & $\beta$    & 1.6                & 2.9   & 1.8-3.4\tnote{1}  & 5.6          & 8.3        \\
 \hline\hline
 \end{tabular}
 \begin{tablenotes}
 \footnotesize
 \item[1] The linewidth of 1uc-FeSe/STO varies with the substrate batches and film growth conditions. The average value is 9.1$\pm$1.5 meV for $\alpha$ mode, and 2.4$\pm$0.9 meV for $\beta$ mode.
 \end{tablenotes}
 \end{threeparttable}
 \end{table*}

\subsubsection{Linewidth broadening compared to the F-K modes of clean STO}

The third feature is that the linewidths ($\Gamma_{ep}$) of F-K phonon modes associated with 1uc-FeSe/STO are larger than those of clean STO(001) as illustrated in Table.\ref{tb1}. The phonon linewidth broadening can be either from EPC or anharmonic phonon-phonon interaction. The phonon-phonon interactions are strongly temperature dependent and can be neglected at low temperature. So the strength of the mode-specific EPC is approximately proportional to the phonon linewidth at 35 K. After the growth of FeSe film, the linewidth of $\alpha$ mode (~ 9.1$\pm$1.5 meV) is broadened comparing to that of clean STO (~ 6.5$\pm$0.7 meV). This linewidth broadening is a signature of EPC enhancement, implying the penetrated F-K phonons do interact with electrons in FeSe films. While the broadening of the $\beta$ mode is less obvious than the $\alpha$ mode, indicating a relatively weaker coupling between electrons and the $\beta$ mode. However, this broadening at q = 0 is currently a challenge for theory\cite{Wangyanprb2017linewidth,Millis2017prb} of interfacial e-ph coupling, rooted in Migdal-Eliashberg-based approaches with dynamics screening.

Additionally thicker FeSe films accompany with larger linewidth, also shown in Table.\ref{tb1}. Since the electric field generated by STO F-K phonons can penetrate into FeSe films, F-K phonons from substrate STO interact with all the electrons in FeSe films. The total amount of electrons in FeSe films increase with increasing thickness, thus the linewidth broadening of the F-K phonon is an additive effect layer by layer with the growth of FeSe films. As a result, thicker FeSe films will always show a larger phonon linewidth than that of thinner films. Electrons in the FeSe layer closest to FeSe/STO interface contribute the largest electron-phonon coupling. This contribution becomes smaller in the layers further from the interface, because the electric field decays exponentially inside the FeSe flims\cite{ShuyuanZhang2016}.

If EPC of F-K phonons play a vital role on the superconductivity of FeSe films, these high energy F-K phonons should also be present in a variety of oxide substrates. Surface phonons on other oxides, such as rutile TiO$_2$(110), BaTiO$_3$(001) and SrTiO$_3$(110)\cite{CaoYanweiSTO110}, are measured and shown in Fig. \ref{fig6}(c). As a ubiquitous characteristic of oxides\cite{FK1965}, all those sample surfaces have F-K phonons with similar energies that are accompanied with strong electric field, independent of the crystal orientation, surface reconstruction, crystal symmetry, or lattice constant. These facts establish the critical role played by oxide F-K phonons on the T$_C$ enhancement at FeSe/oxides interface. Thus the specific F-K phonon energy scale $\sim$ 100 meV and the metal-oxygen chemical bond strength should be vital for the interfacial T$_C$ enhancement.

\subsection{First-principles Calculations and the Magnetic Structure in FeSe films}

Although the F-K phonons from STO substrate play an essential role,
the $T_C$ enhancement in FeSe-derived systems without the substrate
indicate that these modes do not act alone. Magnetic interactions
have been always speculated as one of the most possible candidate
origins of the pairing in FeSe bulk. However, the existence of
magnetic interaction in FeSe/STO is still elusive, due to the
limitation of the experimental techniques to measure the magnetic
ordering of ultrathin films. In this study our first-principles
calculation results show that the AFM correlation in FeSe is
indispensable to quantitatively reproduce the experimental phonon
dispersions in the ultrathin FeSe films.

\begin{figure*}
\begin{center}
\includegraphics[width=0.9\textwidth]{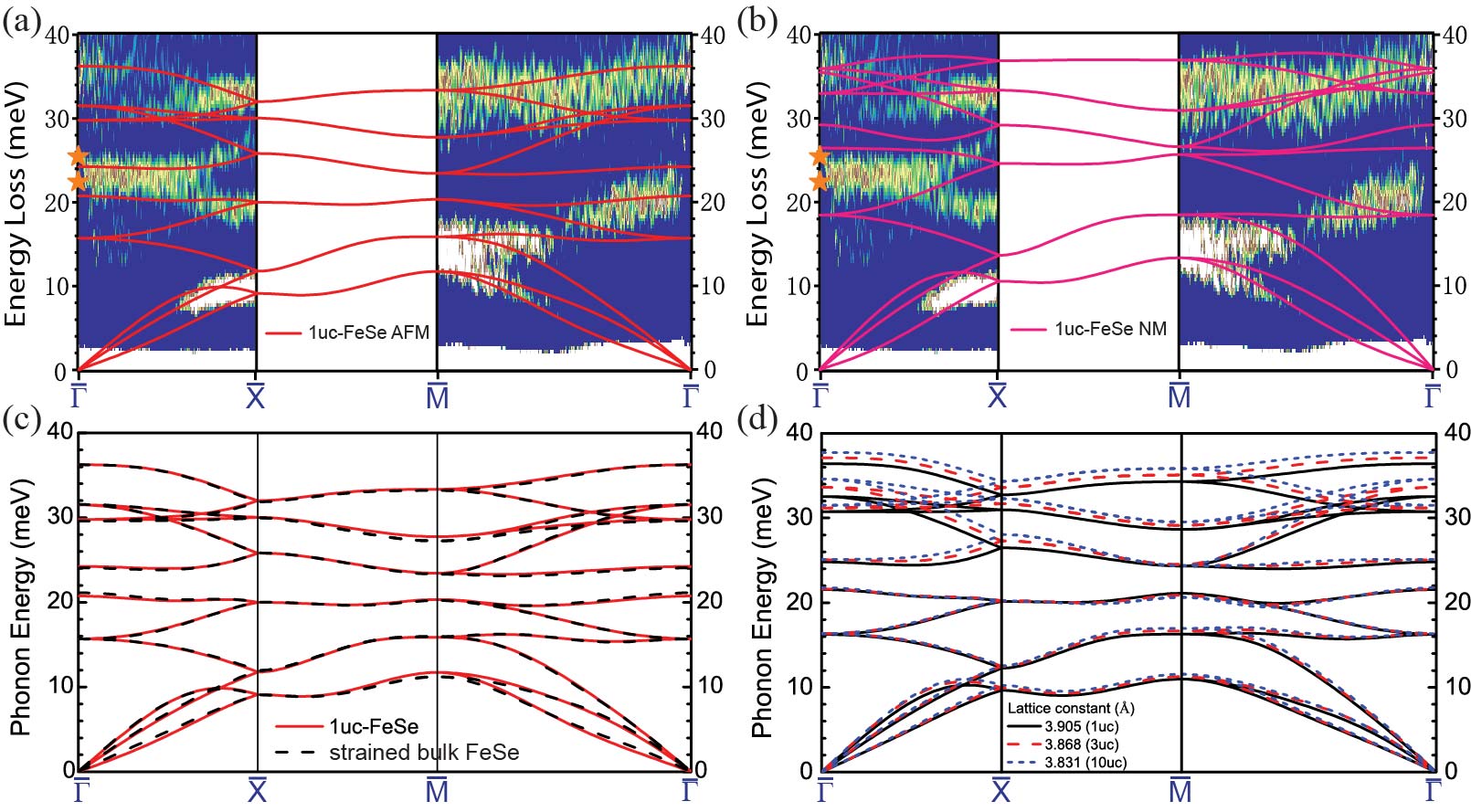}
\caption{\label{fig8} {\small Comparison between experimental data and theoretical calculations for single layer FeSe films, (a) with checkerboard AFM spin configuration on Fe lattice, and (b) without magnetic structure on Fe lattice. Orange stars label the A$_{1g}$ mode (22.6 meV) and B$_{1g}$ mode (25.6 meV) measured by Raman scattering at 7 K\cite{Gnezdilov2013FeSePhonon}. (c) Comparison between theoretical results for single layer and strained bulk FeSe, both with AFM spin configuration on Fe lattice with lattice constant a=3.905 $\AA$ (d) Comparison between theoretical results for bulk FeSe with the lattice constants of 1uc, 3uc and 10uc-FeSe films respectively, both with AFM spin configuration on Fe lattice. Lattice constants used in theoretical calculations are extracted from Ref.\onlinecite{PengRui2014PRLFeSe}.}}
\end{center}
\end{figure*}

First-principles calculations are performed to calculate the
dispersions of the Fe- and Se-derived phonons in the single layer
FeSe film. The technical details of computation have been given in
Section II F. In the calculations, when the AFM spin configuration
on Fe lattice is taken into account, the total energy per Fe is 103
meV lower compared to the non-magnetic configuration. Moreover, the
AFM results exhibit significant phonon energy renormalization and
provide much better consistency with our experimental results than
these from nonmagnetic structure, as shown in Figs. \ref{fig8}(a) and \ref{fig8}(b). In addition, these figures show that the AFM calculations agree
better with Raman scattering at 7 K\cite{Gnezdilov2013FeSePhonon}
compared to the NM calculations. We note that phonon bands 6 and 7
(counting from the low energy bands) at $\overline{\Gamma}$ in the
AFM and NM calculations correspond to the A$_{1g}$(Se) and
B$_{1g}$(Fe) modes respectively. These results verify the existence
of AFM correlations in single layer FeSe films, which implies spin
fluctuation is still important to the superconductivity of single
layer FeSe films, similar to the case in FeSe
bulk\cite{wang2015SpinFluctuation,wang2016SpinFluctuationNC}. As
seen from the comparisons in Figs. \ref{fig8}(c) and \ref{fig8}(d), film
thicknesses and lattice constant cannot significantly modify the energies of FeSe
phonons, which is also consistent with the HREELS
experimental results. The unflappable FeSe phonons for different
film thickness suggest that similar magnetic moments or AFM
correlations are possessed by Fe atoms for various thickness FeSe
films, since the phonon energy directly reflects the AFM ground
state of FeSe as discussed above. As a result, from a similar
strength of AFM correlations, the spin fluctuations are not expected
to be enhanced from the FeSe bulk to thin films. Rather than
directly increasing the AFM correlation in the films to enhance the
superconductivity mediated by the spin fluctuations, the F-K phonons
of STO must take effect by itself or in an indirect way to give a
significantly higher $T_c$, which requires further theoretical and
experimental investigations.

\section{Conclusion}

In summary, although the electronic structure is tuned by STO substrates, lattice dynamics of FeSe films are unaffected. The phonon energy dispersions associated with the Fe and Se atoms are not only temperature-independent but also thickness-independent. In contrast, the F-K phonons of STO substrate are strongly temperature-dependent in line profile, and change drastically with or without FeSe films. Therefore, if there is EPC which could enhance the interfacial superconductivity, it must be the interaction between the F-K modes of the substrate (with electric field penetrating into the film) and electrons in the FeSe film. Coupling to the FeSe derived phonon modes does not increase the superconducting $T_C$. Moreover, combination of the calculations and experimental phonon dispersions strongly suggest the existence of AFM correlations in 1uc-FeSe/STO, which is also verified to be thickness-independent. Since the superconductivity pairing mechanism of bulk FeSe is still elusive, how the F-K phonon modes in the substrates enhance the existing pairing still needs to be elucidated in future studies.

\section*{Acknowledgements}

The work was supported by the National Natural Science Foundation of
China (No. 11634016), the National Key R\&D Program of China (No.
2017YFA0303600). X. Z. was partially supported by the Youth
Innovation Promotion Association of Chinese Academy of Sciences
(CAS), and the Open Research Fund Program of the State Key
Laboratory of Low-Dimensional Quantum Physics. Y. W. was supported
by the U.S. Department of Energy, Office of Basic Energy Sciences,
Materials Sciences and Engineering Division. Part of this work (T.
B.) was conducted at the Center for Nanophase Materials Sciences,
sponsored by the Scientific User Facilities Division (SUFD), Basic
Energy Sciences (BES), DOE, under contract with UT-Battelle. CPU
time was provided in part by resources supported by the University
of Tennessee and Oak Ridge National Laboratory Joint Institute for
Computational Sciences (http://www.jics.utk.edu). This research used
resources of the National Energy Research Scientific Computing
Center, a DOE Office of Science User Facility supported by the
Office of Science of the U.S. DOE under Contract No.
DE-AC02-05CH11231. J. Z. was partially supported by U.S. NSF through
Grant No. DMR 1608865, and the sabbatical program of the Institute
of Physics CAS.

\bibliographystyle{apsrev4-1}

\begin{thebibliography}{75}%
\makeatletter
\providecommand \@ifxundefined [1]{%
 \@ifx{#1\undefined}
}%
\providecommand \@ifnum [1]{%
 \ifnum #1\expandafter \@firstoftwo
 \else \expandafter \@secondoftwo
 \fi
}%
\providecommand \@ifx [1]{%
 \ifx #1\expandafter \@firstoftwo
 \else \expandafter \@secondoftwo
 \fi
}%
\providecommand \natexlab [1]{#1}%
\providecommand \enquote  [1]{``#1''}%
\providecommand \bibnamefont  [1]{#1}%
\providecommand \bibfnamefont [1]{#1}%
\providecommand \citenamefont [1]{#1}%
\providecommand \href@noop [0]{\@secondoftwo}%
\providecommand \href [0]{\begingroup \@sanitize@url \@href}%
\providecommand \@href[1]{\@@startlink{#1}\@@href}%
\providecommand \@@href[1]{\endgroup#1\@@endlink}%
\providecommand \@sanitize@url [0]{\catcode `\\12\catcode `\$12\catcode
  `\&12\catcode `\#12\catcode `\^12\catcode `\_12\catcode `\%12\relax}%
\providecommand \@@startlink[1]{}%
\providecommand \@@endlink[0]{}%
\providecommand \url  [0]{\begingroup\@sanitize@url \@url }%
\providecommand \@url [1]{\endgroup\@href {#1}{\urlprefix }}%
\providecommand \urlprefix  [0]{URL }%
\providecommand \Eprint [0]{\href }%
\providecommand \doibase [0]{http://dx.doi.org/}%
\providecommand \selectlanguage [0]{\@gobble}%
\providecommand \bibinfo  [0]{\@secondoftwo}%
\providecommand \bibfield  [0]{\@secondoftwo}%
\providecommand \translation [1]{[#1]}%
\providecommand \BibitemOpen [0]{}%
\providecommand \bibitemStop [0]{}%
\providecommand \bibitemNoStop [0]{.\EOS\space}%
\providecommand \EOS [0]{\spacefactor3000\relax}%
\providecommand \BibitemShut  [1]{\csname bibitem#1\endcsname}%
\let\auto@bib@innerbib\@empty
\bibitem [{\citenamefont {Wang}\ \emph {et~al.}(2012)\citenamefont {Wang},
  \citenamefont {Li}, \citenamefont {Zhang}, \citenamefont {Zhang},
  \citenamefont {Zhang}, \citenamefont {Li}, \citenamefont {Ding},
  \citenamefont {Ou}, \citenamefont {Deng}, \citenamefont {Chang},
  \citenamefont {Wen}, \citenamefont {Song}, \citenamefont {He}, \citenamefont
  {Jia}, \citenamefont {Ji}, \citenamefont {Wang}, \citenamefont {Wang},
  \citenamefont {Chen}, \citenamefont {Ma},\ and\ \citenamefont
  {Xue}}]{WangQingyan2012-1ucFeSe}%
  \BibitemOpen
  \bibfield  {author} {\bibinfo {author} {\bibfnamefont {Q.-Y.}\ \bibnamefont
  {Wang}}, \bibinfo {author} {\bibfnamefont {Z.}~\bibnamefont {Li}}, \bibinfo
  {author} {\bibfnamefont {W.-H.}\ \bibnamefont {Zhang}}, \bibinfo {author}
  {\bibfnamefont {Z.-C.}\ \bibnamefont {Zhang}}, \bibinfo {author}
  {\bibfnamefont {J.-S.}\ \bibnamefont {Zhang}}, \bibinfo {author}
  {\bibfnamefont {W.}~\bibnamefont {Li}}, \bibinfo {author} {\bibfnamefont
  {H.}~\bibnamefont {Ding}}, \bibinfo {author} {\bibfnamefont {Y.-B.}\
  \bibnamefont {Ou}}, \bibinfo {author} {\bibfnamefont {P.}~\bibnamefont
  {Deng}}, \bibinfo {author} {\bibfnamefont {K.}~\bibnamefont {Chang}},
  \bibinfo {author} {\bibfnamefont {J.}~\bibnamefont {Wen}}, \bibinfo {author}
  {\bibfnamefont {C.-L.}\ \bibnamefont {Song}}, \bibinfo {author}
  {\bibfnamefont {K.}~\bibnamefont {He}}, \bibinfo {author} {\bibfnamefont
  {J.-F.}\ \bibnamefont {Jia}}, \bibinfo {author} {\bibfnamefont {S.-H.}\
  \bibnamefont {Ji}}, \bibinfo {author} {\bibfnamefont {Y.-Y.}\ \bibnamefont
  {Wang}}, \bibinfo {author} {\bibfnamefont {L.-L.}\ \bibnamefont {Wang}},
  \bibinfo {author} {\bibfnamefont {X.}~\bibnamefont {Chen}}, \bibinfo {author}
  {\bibfnamefont {X.-C.}\ \bibnamefont {Ma}}, \ and\ \bibinfo {author}
  {\bibfnamefont {Q.-K.}\ \bibnamefont {Xue}},\ }\href
  {http://stacks.iop.org/0256-307X/29/i=3/a=037402} {\bibfield  {journal}
  {\bibinfo  {journal} {Chinese Physics Letters}\ }\textbf {\bibinfo {volume}
  {29}},\ \bibinfo {pages} {037402} (\bibinfo {year} {2012})}\BibitemShut
  {NoStop}%
\bibitem [{\citenamefont {Zhang}\ \emph
  {et~al.}(2014{\natexlab{a}})\citenamefont {Zhang}, \citenamefont {Sun},
  \citenamefont {Zhang}, \citenamefont {Li}, \citenamefont {Guo}, \citenamefont
  {Zhao}, \citenamefont {Zhang}, \citenamefont {Peng}, \citenamefont {Xing},
  \citenamefont {Wang}, \citenamefont {Fujita}, \citenamefont {Hirata},
  \citenamefont {Li}, \citenamefont {Ding}, \citenamefont {Tang}, \citenamefont
  {Wang}, \citenamefont {Wang}, \citenamefont {He}, \citenamefont {Ji},
  \citenamefont {Chen}, \citenamefont {Wang}, \citenamefont {Xia},
  \citenamefont {Li}, \citenamefont {Wang}, \citenamefont {Wang}, \citenamefont
  {Wang}, \citenamefont {Chen}, \citenamefont {Xue},\ and\ \citenamefont
  {Ma}}]{ZhangWenhao2014-CPL-1ucFeSe}%
  \BibitemOpen
  \bibfield  {author} {\bibinfo {author} {\bibfnamefont {W.-H.}\ \bibnamefont
  {Zhang}}, \bibinfo {author} {\bibfnamefont {Y.}~\bibnamefont {Sun}}, \bibinfo
  {author} {\bibfnamefont {J.-S.}\ \bibnamefont {Zhang}}, \bibinfo {author}
  {\bibfnamefont {F.-S.}\ \bibnamefont {Li}}, \bibinfo {author} {\bibfnamefont
  {M.-H.}\ \bibnamefont {Guo}}, \bibinfo {author} {\bibfnamefont {Y.-F.}\
  \bibnamefont {Zhao}}, \bibinfo {author} {\bibfnamefont {H.-M.}\ \bibnamefont
  {Zhang}}, \bibinfo {author} {\bibfnamefont {J.-P.}\ \bibnamefont {Peng}},
  \bibinfo {author} {\bibfnamefont {Y.}~\bibnamefont {Xing}}, \bibinfo {author}
  {\bibfnamefont {H.-C.}\ \bibnamefont {Wang}}, \bibinfo {author}
  {\bibfnamefont {T.}~\bibnamefont {Fujita}}, \bibinfo {author} {\bibfnamefont
  {A.}~\bibnamefont {Hirata}}, \bibinfo {author} {\bibfnamefont
  {Z.}~\bibnamefont {Li}}, \bibinfo {author} {\bibfnamefont {H.}~\bibnamefont
  {Ding}}, \bibinfo {author} {\bibfnamefont {C.-J.}\ \bibnamefont {Tang}},
  \bibinfo {author} {\bibfnamefont {M.}~\bibnamefont {Wang}}, \bibinfo {author}
  {\bibfnamefont {Q.-Y.}\ \bibnamefont {Wang}}, \bibinfo {author}
  {\bibfnamefont {K.}~\bibnamefont {He}}, \bibinfo {author} {\bibfnamefont
  {S.-H.}\ \bibnamefont {Ji}}, \bibinfo {author} {\bibfnamefont
  {X.}~\bibnamefont {Chen}}, \bibinfo {author} {\bibfnamefont {J.-F.}\
  \bibnamefont {Wang}}, \bibinfo {author} {\bibfnamefont {Z.-C.}\ \bibnamefont
  {Xia}}, \bibinfo {author} {\bibfnamefont {L.}~\bibnamefont {Li}}, \bibinfo
  {author} {\bibfnamefont {Y.-Y.}\ \bibnamefont {Wang}}, \bibinfo {author}
  {\bibfnamefont {J.}~\bibnamefont {Wang}}, \bibinfo {author} {\bibfnamefont
  {L.-L.}\ \bibnamefont {Wang}}, \bibinfo {author} {\bibfnamefont {M.-W.}\
  \bibnamefont {Chen}}, \bibinfo {author} {\bibfnamefont {Q.-K.}\ \bibnamefont
  {Xue}}, \ and\ \bibinfo {author} {\bibfnamefont {X.-C.}\ \bibnamefont {Ma}},\
  }\href {http://stacks.iop.org/0256-307X/31/i=1/a=017401} {\bibfield
  {journal} {\bibinfo  {journal} {Chinese Physics Letters}\ }\textbf {\bibinfo
  {volume} {31}},\ \bibinfo {pages} {017401} (\bibinfo {year}
  {2014}{\natexlab{a}})}\BibitemShut {NoStop}%
\bibitem [{\citenamefont {Zhang}\ \emph
  {et~al.}(2014{\natexlab{b}})\citenamefont {Zhang}, \citenamefont {Li},
  \citenamefont {Li}, \citenamefont {Zhang}, \citenamefont {Peng},
  \citenamefont {Tang}, \citenamefont {Wang}, \citenamefont {He}, \citenamefont
  {Chen}, \citenamefont {Wang}, \citenamefont {Ma},\ and\ \citenamefont
  {Xue}}]{ZhangWenhao2014-PhysRevB-1ucFeSe}%
  \BibitemOpen
  \bibfield  {author} {\bibinfo {author} {\bibfnamefont {W.}~\bibnamefont
  {Zhang}}, \bibinfo {author} {\bibfnamefont {Z.}~\bibnamefont {Li}}, \bibinfo
  {author} {\bibfnamefont {F.}~\bibnamefont {Li}}, \bibinfo {author}
  {\bibfnamefont {H.}~\bibnamefont {Zhang}}, \bibinfo {author} {\bibfnamefont
  {J.}~\bibnamefont {Peng}}, \bibinfo {author} {\bibfnamefont {C.}~\bibnamefont
  {Tang}}, \bibinfo {author} {\bibfnamefont {Q.}~\bibnamefont {Wang}}, \bibinfo
  {author} {\bibfnamefont {K.}~\bibnamefont {He}}, \bibinfo {author}
  {\bibfnamefont {X.}~\bibnamefont {Chen}}, \bibinfo {author} {\bibfnamefont
  {L.}~\bibnamefont {Wang}}, \bibinfo {author} {\bibfnamefont {X.}~\bibnamefont
  {Ma}}, \ and\ \bibinfo {author} {\bibfnamefont {Q.-K.}\ \bibnamefont {Xue}},\
  }\href {\doibase 10.1103/PhysRevB.89.060506} {\bibfield  {journal} {\bibinfo
  {journal} {Phys. Rev. B}\ }\textbf {\bibinfo {volume} {89}},\ \bibinfo
  {pages} {060506} (\bibinfo {year} {2014}{\natexlab{b}})}\BibitemShut
  {NoStop}%
\bibitem [{\citenamefont {Liu}\ \emph {et~al.}(2012)\citenamefont {Liu},
  \citenamefont {Zhang}, \citenamefont {Mou}, \citenamefont {He}, \citenamefont
  {Ou}, \citenamefont {Wang}, \citenamefont {Li}, \citenamefont {Wang},
  \citenamefont {Zhao}, \citenamefont {He} \emph {et~al.}}]{liuDefa2012FeSe}%
  \BibitemOpen
  \bibfield  {author} {\bibinfo {author} {\bibfnamefont {D.}~\bibnamefont
  {Liu}}, \bibinfo {author} {\bibfnamefont {W.}~\bibnamefont {Zhang}}, \bibinfo
  {author} {\bibfnamefont {D.}~\bibnamefont {Mou}}, \bibinfo {author}
  {\bibfnamefont {J.}~\bibnamefont {He}}, \bibinfo {author} {\bibfnamefont
  {Y.-B.}\ \bibnamefont {Ou}}, \bibinfo {author} {\bibfnamefont {Q.-Y.}\
  \bibnamefont {Wang}}, \bibinfo {author} {\bibfnamefont {Z.}~\bibnamefont
  {Li}}, \bibinfo {author} {\bibfnamefont {L.}~\bibnamefont {Wang}}, \bibinfo
  {author} {\bibfnamefont {L.}~\bibnamefont {Zhao}}, \bibinfo {author}
  {\bibfnamefont {S.}~\bibnamefont {He}},  \emph {et~al.},\ }\href
  {http://www.nature.com/ncomms/journal/v3/n7/abs/ncomms1946.html} {\bibfield
  {journal} {\bibinfo  {journal} {Nature communications}\ }\textbf {\bibinfo
  {volume} {3}},\ \bibinfo {pages} {931} (\bibinfo {year} {2012})}\BibitemShut
  {NoStop}%
\bibitem [{\citenamefont {Lee}\ \emph {et~al.}(2014)\citenamefont {Lee},
  \citenamefont {Schmitt}, \citenamefont {Moore}, \citenamefont {Johnston},
  \citenamefont {Cui}, \citenamefont {Li}, \citenamefont {Yi}, \citenamefont
  {Liu}, \citenamefont {Hashimoto}, \citenamefont {Zhang} \emph
  {et~al.}}]{LeeJJ2014FeSereplica}%
  \BibitemOpen
  \bibfield  {author} {\bibinfo {author} {\bibfnamefont {J.}~\bibnamefont
  {Lee}}, \bibinfo {author} {\bibfnamefont {F.}~\bibnamefont {Schmitt}},
  \bibinfo {author} {\bibfnamefont {R.}~\bibnamefont {Moore}}, \bibinfo
  {author} {\bibfnamefont {S.}~\bibnamefont {Johnston}}, \bibinfo {author}
  {\bibfnamefont {Y.-T.}\ \bibnamefont {Cui}}, \bibinfo {author} {\bibfnamefont
  {W.}~\bibnamefont {Li}}, \bibinfo {author} {\bibfnamefont {M.}~\bibnamefont
  {Yi}}, \bibinfo {author} {\bibfnamefont {Z.}~\bibnamefont {Liu}}, \bibinfo
  {author} {\bibfnamefont {M.}~\bibnamefont {Hashimoto}}, \bibinfo {author}
  {\bibfnamefont {Y.}~\bibnamefont {Zhang}},  \emph {et~al.},\ }\href {\doibase
  10.1038/nature13894} {\bibfield  {journal} {\bibinfo  {journal} {Nature}\
  }\textbf {\bibinfo {volume} {515}},\ \bibinfo {pages} {245} (\bibinfo {year}
  {2014})}\BibitemShut {NoStop}%
\bibitem [{\citenamefont {Tan}\ \emph {et~al.}(2013)\citenamefont {Tan},
  \citenamefont {Zhang}, \citenamefont {Xia}, \citenamefont {Ye}, \citenamefont
  {Chen}, \citenamefont {Xie}, \citenamefont {Peng}, \citenamefont {Xu},
  \citenamefont {Fan}, \citenamefont {Xu} \emph {et~al.}}]{TanShiyong2013FeSe}%
  \BibitemOpen
  \bibfield  {author} {\bibinfo {author} {\bibfnamefont {S.}~\bibnamefont
  {Tan}}, \bibinfo {author} {\bibfnamefont {Y.}~\bibnamefont {Zhang}}, \bibinfo
  {author} {\bibfnamefont {M.}~\bibnamefont {Xia}}, \bibinfo {author}
  {\bibfnamefont {Z.}~\bibnamefont {Ye}}, \bibinfo {author} {\bibfnamefont
  {F.}~\bibnamefont {Chen}}, \bibinfo {author} {\bibfnamefont {X.}~\bibnamefont
  {Xie}}, \bibinfo {author} {\bibfnamefont {R.}~\bibnamefont {Peng}}, \bibinfo
  {author} {\bibfnamefont {D.}~\bibnamefont {Xu}}, \bibinfo {author}
  {\bibfnamefont {Q.}~\bibnamefont {Fan}}, \bibinfo {author} {\bibfnamefont
  {H.}~\bibnamefont {Xu}},  \emph {et~al.},\ }\href {\doibase 0.1038/nmat3654}
  {\bibfield  {journal} {\bibinfo  {journal} {Nature materials}\ }\textbf
  {\bibinfo {volume} {12}},\ \bibinfo {pages} {634} (\bibinfo {year}
  {2013})}\BibitemShut {NoStop}%
\bibitem [{\citenamefont {Ge}\ \emph {et~al.}(2015)\citenamefont {Ge},
  \citenamefont {Liu}, \citenamefont {Liu}, \citenamefont {Gao}, \citenamefont
  {Qian}, \citenamefont {Xue}, \citenamefont {Liu},\ and\ \citenamefont
  {Jia}}]{GeJianfeng2015FeSe}%
  \BibitemOpen
  \bibfield  {author} {\bibinfo {author} {\bibfnamefont {J.-F.}\ \bibnamefont
  {Ge}}, \bibinfo {author} {\bibfnamefont {Z.-L.}\ \bibnamefont {Liu}},
  \bibinfo {author} {\bibfnamefont {C.}~\bibnamefont {Liu}}, \bibinfo {author}
  {\bibfnamefont {C.-L.}\ \bibnamefont {Gao}}, \bibinfo {author} {\bibfnamefont
  {D.}~\bibnamefont {Qian}}, \bibinfo {author} {\bibfnamefont {Q.-K.}\
  \bibnamefont {Xue}}, \bibinfo {author} {\bibfnamefont {Y.}~\bibnamefont
  {Liu}}, \ and\ \bibinfo {author} {\bibfnamefont {J.-F.}\ \bibnamefont
  {Jia}},\ }\href
  {http://www.nature.com/nmat/journal/v14/n3/full/nmat4153.html} {\bibfield
  {journal} {\bibinfo  {journal} {Nature materials}\ }\textbf {\bibinfo
  {volume} {14}},\ \bibinfo {pages} {285} (\bibinfo {year} {2015})}\BibitemShut
  {NoStop}%
\bibitem [{\citenamefont {Hsu}\ \emph {et~al.}(2008)\citenamefont {Hsu},
  \citenamefont {Luo}, \citenamefont {Yeh}, \citenamefont {Chen}, \citenamefont
  {Huang}, \citenamefont {Wu}, \citenamefont {Lee}, \citenamefont {Huang},
  \citenamefont {Chu}, \citenamefont {Yan} \emph {et~al.}}]{hsu2008FeSe}%
  \BibitemOpen
  \bibfield  {author} {\bibinfo {author} {\bibfnamefont {F.-C.}\ \bibnamefont
  {Hsu}}, \bibinfo {author} {\bibfnamefont {J.-Y.}\ \bibnamefont {Luo}},
  \bibinfo {author} {\bibfnamefont {K.-W.}\ \bibnamefont {Yeh}}, \bibinfo
  {author} {\bibfnamefont {T.-K.}\ \bibnamefont {Chen}}, \bibinfo {author}
  {\bibfnamefont {T.-W.}\ \bibnamefont {Huang}}, \bibinfo {author}
  {\bibfnamefont {P.~M.}\ \bibnamefont {Wu}}, \bibinfo {author} {\bibfnamefont
  {Y.-C.}\ \bibnamefont {Lee}}, \bibinfo {author} {\bibfnamefont {Y.-L.}\
  \bibnamefont {Huang}}, \bibinfo {author} {\bibfnamefont {Y.-Y.}\ \bibnamefont
  {Chu}}, \bibinfo {author} {\bibfnamefont {D.-C.}\ \bibnamefont {Yan}},  \emph
  {et~al.},\ }\href {http://www.pnas.org/content/105/38/14262.short} {\bibfield
   {journal} {\bibinfo  {journal} {Proceedings of the National Academy of
  Sciences}\ }\textbf {\bibinfo {volume} {105}},\ \bibinfo {pages} {14262}
  (\bibinfo {year} {2008})}\BibitemShut {NoStop}%
\bibitem [{\citenamefont {Lee}(2015)}]{LiDunghai2015CPB}%
  \BibitemOpen
  \bibfield  {author} {\bibinfo {author} {\bibfnamefont {D.-H.}\ \bibnamefont
  {Lee}},\ }\href {http://stacks.iop.org/1674-1056/24/i=11/a=117405} {\bibfield
   {journal} {\bibinfo  {journal} {Chinese Physics B}\ }\textbf {\bibinfo
  {volume} {24}},\ \bibinfo {pages} {117405} (\bibinfo {year}
  {2015})}\BibitemShut {NoStop}%
\bibitem [{\citenamefont {Bozovic}\ and\ \citenamefont
  {Ahn}(2014)}]{bozovic2014NP}%
  \BibitemOpen
  \bibfield  {author} {\bibinfo {author} {\bibfnamefont {I.}~\bibnamefont
  {Bozovic}}\ and\ \bibinfo {author} {\bibfnamefont {C.}~\bibnamefont {Ahn}},\
  }\href {\doibase doi:10.1038/nphys3177} {\bibfield  {journal} {\bibinfo
  {journal} {Nature Physics}\ }\textbf {\bibinfo {volume} {10}},\ \bibinfo
  {pages} {892} (\bibinfo {year} {2014})}\BibitemShut {NoStop}%
\bibitem [{\citenamefont {Huang}\ and\ \citenamefont
  {Hoffman}(2017)}]{Hoffman2017monolayerFeSeReview}%
  \BibitemOpen
  \bibfield  {author} {\bibinfo {author} {\bibfnamefont {D.}~\bibnamefont
  {Huang}}\ and\ \bibinfo {author} {\bibfnamefont {J.~E.}\ \bibnamefont
  {Hoffman}},\ }\href
  {https://doi.org/10.1146/annurev-conmatphys-031016-025242} {\bibfield
  {journal} {\bibinfo  {journal} {Annual Review of Condensed Matter Physics}\
  }\textbf {\bibinfo {volume} {8}},\ \bibinfo {pages} {311} (\bibinfo {year}
  {2017})}\BibitemShut {NoStop}%
\bibitem [{\citenamefont {Wang}\ \emph
  {et~al.}(2017{\natexlab{a}})\citenamefont {Wang}, \citenamefont {Liu},
  \citenamefont {Liu},\ and\ \citenamefont {Wang}}]{wang2017FeSeReview}%
  \BibitemOpen
  \bibfield  {author} {\bibinfo {author} {\bibfnamefont {Z.}~\bibnamefont
  {Wang}}, \bibinfo {author} {\bibfnamefont {C.}~\bibnamefont {Liu}}, \bibinfo
  {author} {\bibfnamefont {Y.}~\bibnamefont {Liu}}, \ and\ \bibinfo {author}
  {\bibfnamefont {J.}~\bibnamefont {Wang}},\ }\href
  {http://stacks.iop.org/0953-8984/29/i=15/a=153001} {\bibfield  {journal}
  {\bibinfo  {journal} {Journal of Physics: Condensed Matter}\ }\textbf
  {\bibinfo {volume} {29}},\ \bibinfo {pages} {153001} (\bibinfo {year}
  {2017}{\natexlab{a}})}\BibitemShut {NoStop}%
\bibitem [{\citenamefont {Wang}\ \emph
  {et~al.}(2016{\natexlab{a}})\citenamefont {Wang}, \citenamefont {Ma},\ and\
  \citenamefont {Xue}}]{Xue2016FeSeReview}%
  \BibitemOpen
  \bibfield  {author} {\bibinfo {author} {\bibfnamefont {L.}~\bibnamefont
  {Wang}}, \bibinfo {author} {\bibfnamefont {X.}~\bibnamefont {Ma}}, \ and\
  \bibinfo {author} {\bibfnamefont {Q.-K.}\ \bibnamefont {Xue}},\ }\href
  {http://stacks.iop.org/0953-2048/29/i=12/a=123001} {\bibfield  {journal}
  {\bibinfo  {journal} {Superconductor Science and Technology}\ }\textbf
  {\bibinfo {volume} {29}},\ \bibinfo {pages} {123001} (\bibinfo {year}
  {2016}{\natexlab{a}})}\BibitemShut {NoStop}%
\bibitem [{\citenamefont {Zhao}\ \emph {et~al.}(2017)\citenamefont {Zhao},
  \citenamefont {Li}, \citenamefont {Chang}, \citenamefont {Jiang},
  \citenamefont {Wu}, \citenamefont {Liu}, \citenamefont {Zhu}, \citenamefont
  {Moodera},\ and\ \citenamefont {Chan}}]{Chen2017TEM}%
  \BibitemOpen
  \bibfield  {author} {\bibinfo {author} {\bibfnamefont {W.}~\bibnamefont
  {Zhao}}, \bibinfo {author} {\bibfnamefont {M.}~\bibnamefont {Li}}, \bibinfo
  {author} {\bibfnamefont {C.-Z.}\ \bibnamefont {Chang}}, \bibinfo {author}
  {\bibfnamefont {J.}~\bibnamefont {Jiang}}, \bibinfo {author} {\bibfnamefont
  {L.}~\bibnamefont {Wu}}, \bibinfo {author} {\bibfnamefont {C.}~\bibnamefont
  {Liu}}, \bibinfo {author} {\bibfnamefont {Y.}~\bibnamefont {Zhu}}, \bibinfo
  {author} {\bibfnamefont {J.~S.}\ \bibnamefont {Moodera}}, \ and\ \bibinfo
  {author} {\bibfnamefont {M.~H.}\ \bibnamefont {Chan}},\ }\href
  {https://arxiv.org/abs/1701.03678} {\bibfield  {journal} {\bibinfo  {journal}
  {arXiv preprint arXiv:1701.03678}\ } (\bibinfo {year} {2017})}\BibitemShut
  {NoStop}%
\bibitem [{\citenamefont {He}\ \emph {et~al.}(2013)\citenamefont {He},
  \citenamefont {He}, \citenamefont {Zhang}, \citenamefont {Zhao},
  \citenamefont {Liu}, \citenamefont {Liu}, \citenamefont {Mou}, \citenamefont
  {Ou}, \citenamefont {Wang}, \citenamefont {Li} \emph
  {et~al.}}]{HeShaolong2013FeSe}%
  \BibitemOpen
  \bibfield  {author} {\bibinfo {author} {\bibfnamefont {S.}~\bibnamefont
  {He}}, \bibinfo {author} {\bibfnamefont {J.}~\bibnamefont {He}}, \bibinfo
  {author} {\bibfnamefont {W.}~\bibnamefont {Zhang}}, \bibinfo {author}
  {\bibfnamefont {L.}~\bibnamefont {Zhao}}, \bibinfo {author} {\bibfnamefont
  {D.}~\bibnamefont {Liu}}, \bibinfo {author} {\bibfnamefont {X.}~\bibnamefont
  {Liu}}, \bibinfo {author} {\bibfnamefont {D.}~\bibnamefont {Mou}}, \bibinfo
  {author} {\bibfnamefont {Y.-B.}\ \bibnamefont {Ou}}, \bibinfo {author}
  {\bibfnamefont {Q.-Y.}\ \bibnamefont {Wang}}, \bibinfo {author}
  {\bibfnamefont {Z.}~\bibnamefont {Li}},  \emph {et~al.},\ }\href {\doibase
  doi:10.1038/nmat3648} {\bibfield  {journal} {\bibinfo  {journal} {Nature
  materials}\ }\textbf {\bibinfo {volume} {12}},\ \bibinfo {pages} {605}
  (\bibinfo {year} {2013})}\BibitemShut {NoStop}%
\bibitem [{\citenamefont {He}\ \emph {et~al.}(2014)\citenamefont {He},
  \citenamefont {Liu}, \citenamefont {Zhang}, \citenamefont {Zhao},
  \citenamefont {Liu}, \citenamefont {He}, \citenamefont {Mou}, \citenamefont
  {Li}, \citenamefont {Tang}, \citenamefont {Li} \emph
  {et~al.}}]{HeJunfeng2014FeSe}%
  \BibitemOpen
  \bibfield  {author} {\bibinfo {author} {\bibfnamefont {J.}~\bibnamefont
  {He}}, \bibinfo {author} {\bibfnamefont {X.}~\bibnamefont {Liu}}, \bibinfo
  {author} {\bibfnamefont {W.}~\bibnamefont {Zhang}}, \bibinfo {author}
  {\bibfnamefont {L.}~\bibnamefont {Zhao}}, \bibinfo {author} {\bibfnamefont
  {D.}~\bibnamefont {Liu}}, \bibinfo {author} {\bibfnamefont {S.}~\bibnamefont
  {He}}, \bibinfo {author} {\bibfnamefont {D.}~\bibnamefont {Mou}}, \bibinfo
  {author} {\bibfnamefont {F.}~\bibnamefont {Li}}, \bibinfo {author}
  {\bibfnamefont {C.}~\bibnamefont {Tang}}, \bibinfo {author} {\bibfnamefont
  {Z.}~\bibnamefont {Li}},  \emph {et~al.},\ }\href {\doibase
  10.1073/pnas.1414094112} {\bibfield  {journal} {\bibinfo  {journal}
  {Proceedings of the National Academy of Sciences}\ }\textbf {\bibinfo
  {volume} {111}},\ \bibinfo {pages} {18501} (\bibinfo {year}
  {2014})}\BibitemShut {NoStop}%
\bibitem [{\citenamefont {Lu}\ \emph {et~al.}(2015)\citenamefont {Lu},
  \citenamefont {Wang}, \citenamefont {Wu}, \citenamefont {Wu}, \citenamefont
  {Zhao}, \citenamefont {Zeng}, \citenamefont {Luo}, \citenamefont {Wu},
  \citenamefont {Bao}, \citenamefont {Zhang}, \citenamefont {Huang},
  \citenamefont {Huang},\ and\ \citenamefont
  {Chen}}]{Chenxianhui2015LiFeOHFeSe}%
  \BibitemOpen
  \bibfield  {author} {\bibinfo {author} {\bibfnamefont {X.}~\bibnamefont
  {Lu}}, \bibinfo {author} {\bibfnamefont {N.}~\bibnamefont {Wang}}, \bibinfo
  {author} {\bibfnamefont {H.}~\bibnamefont {Wu}}, \bibinfo {author}
  {\bibfnamefont {Y.}~\bibnamefont {Wu}}, \bibinfo {author} {\bibfnamefont
  {D.}~\bibnamefont {Zhao}}, \bibinfo {author} {\bibfnamefont {X.}~\bibnamefont
  {Zeng}}, \bibinfo {author} {\bibfnamefont {X.}~\bibnamefont {Luo}}, \bibinfo
  {author} {\bibfnamefont {T.}~\bibnamefont {Wu}}, \bibinfo {author}
  {\bibfnamefont {W.}~\bibnamefont {Bao}}, \bibinfo {author} {\bibfnamefont
  {G.}~\bibnamefont {Zhang}}, \bibinfo {author} {\bibfnamefont
  {F.}~\bibnamefont {Huang}}, \bibinfo {author} {\bibfnamefont
  {Q.}~\bibnamefont {Huang}}, \ and\ \bibinfo {author} {\bibfnamefont
  {X.}~\bibnamefont {Chen}},\ }\href {\doibase doi:10.1038/nmat4155} {\bibfield
   {journal} {\bibinfo  {journal} {Nature Materials}\ }\textbf {\bibinfo
  {volume} {14}},\ \bibinfo {pages} {325} (\bibinfo {year} {2015})}\BibitemShut
  {NoStop}%
\bibitem [{\citenamefont {Zhao}\ \emph {et~al.}(2016)\citenamefont {Zhao},
  \citenamefont {Liang}, \citenamefont {Yuan}, \citenamefont {Hu},
  \citenamefont {Liu}, \citenamefont {Huang}, \citenamefont {He}, \citenamefont
  {Shen}, \citenamefont {Xu}, \citenamefont {Liu}, \citenamefont {Yu},
  \citenamefont {Liu}, \citenamefont {Zhou}, \citenamefont {Huang},
  \citenamefont {Dong}, \citenamefont {Zhou}, \citenamefont {Liu},
  \citenamefont {Lu}, \citenamefont {Zhao}, \citenamefont {Chen}, \citenamefont
  {Xu},\ and\ \citenamefont {Zhou}}]{ZhaoLin2015LiFeOHFeSe}%
  \BibitemOpen
  \bibfield  {author} {\bibinfo {author} {\bibfnamefont {L.}~\bibnamefont
  {Zhao}}, \bibinfo {author} {\bibfnamefont {A.}~\bibnamefont {Liang}},
  \bibinfo {author} {\bibfnamefont {D.}~\bibnamefont {Yuan}}, \bibinfo {author}
  {\bibfnamefont {Y.}~\bibnamefont {Hu}}, \bibinfo {author} {\bibfnamefont
  {D.}~\bibnamefont {Liu}}, \bibinfo {author} {\bibfnamefont {J.}~\bibnamefont
  {Huang}}, \bibinfo {author} {\bibfnamefont {S.}~\bibnamefont {He}}, \bibinfo
  {author} {\bibfnamefont {B.}~\bibnamefont {Shen}}, \bibinfo {author}
  {\bibfnamefont {Y.}~\bibnamefont {Xu}}, \bibinfo {author} {\bibfnamefont
  {X.}~\bibnamefont {Liu}}, \bibinfo {author} {\bibfnamefont {L.}~\bibnamefont
  {Yu}}, \bibinfo {author} {\bibfnamefont {G.}~\bibnamefont {Liu}}, \bibinfo
  {author} {\bibfnamefont {H.}~\bibnamefont {Zhou}}, \bibinfo {author}
  {\bibfnamefont {Y.}~\bibnamefont {Huang}}, \bibinfo {author} {\bibfnamefont
  {X.}~\bibnamefont {Dong}}, \bibinfo {author} {\bibfnamefont {F.}~\bibnamefont
  {Zhou}}, \bibinfo {author} {\bibfnamefont {K.}~\bibnamefont {Liu}}, \bibinfo
  {author} {\bibfnamefont {Z.}~\bibnamefont {Lu}}, \bibinfo {author}
  {\bibfnamefont {Z.}~\bibnamefont {Zhao}}, \bibinfo {author} {\bibfnamefont
  {C.}~\bibnamefont {Chen}}, \bibinfo {author} {\bibfnamefont {Z.}~\bibnamefont
  {Xu}}, \ and\ \bibinfo {author} {\bibfnamefont {X.~J.}\ \bibnamefont
  {Zhou}},\ }\href {\doibase 10.1038/ncomms10608} {\bibfield  {journal}
  {\bibinfo  {journal} {Nature Communications}\ }\textbf {\bibinfo {volume}
  {7}},\ \bibinfo {pages} {10608} (\bibinfo {year} {2016})}\BibitemShut
  {NoStop}%
\bibitem [{\citenamefont {Seo}\ \emph {et~al.}(2016)\citenamefont {Seo},
  \citenamefont {Kim}, \citenamefont {Kim}, \citenamefont {Jeong},
  \citenamefont {Ok}, \citenamefont {Kim}, \citenamefont {Denlinger},
  \citenamefont {Mo}, \citenamefont {Kim},\ and\ \citenamefont
  {Kim}}]{Seo2015KdopedFeSe}%
  \BibitemOpen
  \bibfield  {author} {\bibinfo {author} {\bibfnamefont {J.}~\bibnamefont
  {Seo}}, \bibinfo {author} {\bibfnamefont {B.}~\bibnamefont {Kim}}, \bibinfo
  {author} {\bibfnamefont {B.}~\bibnamefont {Kim}}, \bibinfo {author}
  {\bibfnamefont {J.}~\bibnamefont {Jeong}}, \bibinfo {author} {\bibfnamefont
  {J.}~\bibnamefont {Ok}}, \bibinfo {author} {\bibfnamefont {J.~S.}\
  \bibnamefont {Kim}}, \bibinfo {author} {\bibfnamefont {J.}~\bibnamefont
  {Denlinger}}, \bibinfo {author} {\bibfnamefont {S.-K.}\ \bibnamefont {Mo}},
  \bibinfo {author} {\bibfnamefont {C.}~\bibnamefont {Kim}}, \ and\ \bibinfo
  {author} {\bibfnamefont {Y.}~\bibnamefont {Kim}},\ }\href
  {http://www.nature.com/ncomms/2016/160406/ncomms11116/full/ncomms11116.html}
  {\bibfield  {journal} {\bibinfo  {journal} {Nature Communications}\ }\textbf
  {\bibinfo {volume} {7}},\ \bibinfo {pages} {11116} (\bibinfo {year}
  {2016})}\BibitemShut {NoStop}%
\bibitem [{\citenamefont {Wen}\ \emph {et~al.}(2016)\citenamefont {Wen},
  \citenamefont {Xu}, \citenamefont {Chen}, \citenamefont {Huang},
  \citenamefont {Lou}, \citenamefont {Pu}, \citenamefont {Song}, \citenamefont
  {Xie}, \citenamefont {Abdel-Hafiez}, \citenamefont {Chareev} \emph
  {et~al.}}]{CHPwen2016NC}%
  \BibitemOpen
  \bibfield  {author} {\bibinfo {author} {\bibfnamefont {C.}~\bibnamefont
  {Wen}}, \bibinfo {author} {\bibfnamefont {H.}~\bibnamefont {Xu}}, \bibinfo
  {author} {\bibfnamefont {C.}~\bibnamefont {Chen}}, \bibinfo {author}
  {\bibfnamefont {Z.}~\bibnamefont {Huang}}, \bibinfo {author} {\bibfnamefont
  {X.}~\bibnamefont {Lou}}, \bibinfo {author} {\bibfnamefont {Y.}~\bibnamefont
  {Pu}}, \bibinfo {author} {\bibfnamefont {Q.}~\bibnamefont {Song}}, \bibinfo
  {author} {\bibfnamefont {B.}~\bibnamefont {Xie}}, \bibinfo {author}
  {\bibfnamefont {M.}~\bibnamefont {Abdel-Hafiez}}, \bibinfo {author}
  {\bibfnamefont {D.}~\bibnamefont {Chareev}},  \emph {et~al.},\ }\href
  {http://www.nature.com/ncomms/2016/160308/ncomms10840/full/ncomms10840.html}
  {\bibfield  {journal} {\bibinfo  {journal} {Nature communications}\ }\textbf
  {\bibinfo {volume} {7}},\ \bibinfo {pages} {10840} (\bibinfo {year}
  {2016})}\BibitemShut {NoStop}%
\bibitem [{\citenamefont {Miyata}\ \emph {et~al.}(2015)\citenamefont {Miyata},
  \citenamefont {Nakayama}, \citenamefont {Sugawara}, \citenamefont {Sato},\
  and\ \citenamefont {Takahashi}}]{Miyata2015KdopedFeSe}%
  \BibitemOpen
  \bibfield  {author} {\bibinfo {author} {\bibfnamefont {Y.}~\bibnamefont
  {Miyata}}, \bibinfo {author} {\bibfnamefont {K.}~\bibnamefont {Nakayama}},
  \bibinfo {author} {\bibfnamefont {K.}~\bibnamefont {Sugawara}}, \bibinfo
  {author} {\bibfnamefont {T.}~\bibnamefont {Sato}}, \ and\ \bibinfo {author}
  {\bibfnamefont {T.}~\bibnamefont {Takahashi}},\ }\href
  {http://www.nature.com/nmat/journal/v14/n8/full/nmat4302.html} {\bibfield
  {journal} {\bibinfo  {journal} {Nature materials}\ }\textbf {\bibinfo
  {volume} {14}},\ \bibinfo {pages} {775} (\bibinfo {year} {2015})}\BibitemShut
  {NoStop}%
\bibitem [{\citenamefont {Song}\ \emph {et~al.}(2016)\citenamefont {Song},
  \citenamefont {Zhang}, \citenamefont {Zhong}, \citenamefont {Hu},
  \citenamefont {Ji}, \citenamefont {Wang}, \citenamefont {He}, \citenamefont
  {Ma},\ and\ \citenamefont {Xue}}]{SongCanli2016KdopedFeSe}%
  \BibitemOpen
  \bibfield  {author} {\bibinfo {author} {\bibfnamefont {C.-L.}\ \bibnamefont
  {Song}}, \bibinfo {author} {\bibfnamefont {H.-M.}\ \bibnamefont {Zhang}},
  \bibinfo {author} {\bibfnamefont {Y.}~\bibnamefont {Zhong}}, \bibinfo
  {author} {\bibfnamefont {X.-P.}\ \bibnamefont {Hu}}, \bibinfo {author}
  {\bibfnamefont {S.-H.}\ \bibnamefont {Ji}}, \bibinfo {author} {\bibfnamefont
  {L.}~\bibnamefont {Wang}}, \bibinfo {author} {\bibfnamefont {K.}~\bibnamefont
  {He}}, \bibinfo {author} {\bibfnamefont {X.-C.}\ \bibnamefont {Ma}}, \ and\
  \bibinfo {author} {\bibfnamefont {Q.-K.}\ \bibnamefont {Xue}},\ }\href
  {\doibase 10.1103/PhysRevLett.116.157001} {\bibfield  {journal} {\bibinfo
  {journal} {Phys. Rev. Lett.}\ }\textbf {\bibinfo {volume} {116}},\ \bibinfo
  {pages} {157001} (\bibinfo {year} {2016})}\BibitemShut {NoStop}%
\bibitem [{\citenamefont {Shiogai}\ \emph {et~al.}(2015)\citenamefont
  {Shiogai}, \citenamefont {Ito}, \citenamefont {Mitsuhashi}, \citenamefont
  {Nojima},\ and\ \citenamefont {Tsukazaki}}]{Shiogai2015gateFeSe}%
  \BibitemOpen
  \bibfield  {author} {\bibinfo {author} {\bibfnamefont {J.}~\bibnamefont
  {Shiogai}}, \bibinfo {author} {\bibfnamefont {Y.}~\bibnamefont {Ito}},
  \bibinfo {author} {\bibfnamefont {T.}~\bibnamefont {Mitsuhashi}}, \bibinfo
  {author} {\bibfnamefont {T.}~\bibnamefont {Nojima}}, \ and\ \bibinfo {author}
  {\bibfnamefont {A.}~\bibnamefont {Tsukazaki}},\ }\href
  {http://www.nature.com/nphys/journal/v12/n1/full/nphys3530.html} {\bibfield
  {journal} {\bibinfo  {journal} {Nature Physics}\ }\textbf {\bibinfo {volume}
  {12}},\ \bibinfo {pages} {42} (\bibinfo {year} {2015})}\BibitemShut {NoStop}%
\bibitem [{\citenamefont {Lei}\ \emph {et~al.}(2016)\citenamefont {Lei},
  \citenamefont {Cui}, \citenamefont {Xiang}, \citenamefont {Shang},
  \citenamefont {Wang}, \citenamefont {Ye}, \citenamefont {Luo}, \citenamefont
  {Wu}, \citenamefont {Sun},\ and\ \citenamefont {Chen}}]{Lei2015gateFeSe}%
  \BibitemOpen
  \bibfield  {author} {\bibinfo {author} {\bibfnamefont {B.}~\bibnamefont
  {Lei}}, \bibinfo {author} {\bibfnamefont {J.~H.}\ \bibnamefont {Cui}},
  \bibinfo {author} {\bibfnamefont {Z.~J.}\ \bibnamefont {Xiang}}, \bibinfo
  {author} {\bibfnamefont {C.}~\bibnamefont {Shang}}, \bibinfo {author}
  {\bibfnamefont {N.~Z.}\ \bibnamefont {Wang}}, \bibinfo {author}
  {\bibfnamefont {G.~J.}\ \bibnamefont {Ye}}, \bibinfo {author} {\bibfnamefont
  {X.~G.}\ \bibnamefont {Luo}}, \bibinfo {author} {\bibfnamefont
  {T.}~\bibnamefont {Wu}}, \bibinfo {author} {\bibfnamefont {Z.}~\bibnamefont
  {Sun}}, \ and\ \bibinfo {author} {\bibfnamefont {X.~H.}\ \bibnamefont
  {Chen}},\ }\href {\doibase 10.1103/PhysRevLett.116.077002} {\bibfield
  {journal} {\bibinfo  {journal} {Phys. Rev. Lett.}\ }\textbf {\bibinfo
  {volume} {116}},\ \bibinfo {pages} {077002} (\bibinfo {year}
  {2016})}\BibitemShut {NoStop}%
\bibitem [{\citenamefont {Hanzawa}\ \emph {et~al.}(2016)\citenamefont
  {Hanzawa}, \citenamefont {Sato}, \citenamefont {Hiramatsu}, \citenamefont
  {Kamiya},\ and\ \citenamefont {Hosono}}]{hanzawa2015gateFeSe}%
  \BibitemOpen
  \bibfield  {author} {\bibinfo {author} {\bibfnamefont {K.}~\bibnamefont
  {Hanzawa}}, \bibinfo {author} {\bibfnamefont {H.}~\bibnamefont {Sato}},
  \bibinfo {author} {\bibfnamefont {H.}~\bibnamefont {Hiramatsu}}, \bibinfo
  {author} {\bibfnamefont {T.}~\bibnamefont {Kamiya}}, \ and\ \bibinfo {author}
  {\bibfnamefont {H.}~\bibnamefont {Hosono}},\ }\href {\doibase
  10.1073/pnas.1520810113} {\bibfield  {journal} {\bibinfo  {journal}
  {Proceedings of the National Academy of Sciences}\ }\textbf {\bibinfo
  {volume} {113}},\ \bibinfo {pages} {3986} (\bibinfo {year}
  {2016})}\BibitemShut {NoStop}%
\bibitem [{\citenamefont {Peng}\ \emph
  {et~al.}(2014{\natexlab{a}})\citenamefont {Peng}, \citenamefont {Shen},
  \citenamefont {Xie}, \citenamefont {Xu}, \citenamefont {Tan}, \citenamefont
  {Xia}, \citenamefont {Zhang}, \citenamefont {Cao}, \citenamefont {Gong},
  \citenamefont {Hu}, \citenamefont {Xie},\ and\ \citenamefont
  {Feng}}]{PengRui2014PRLFeSe}%
  \BibitemOpen
  \bibfield  {author} {\bibinfo {author} {\bibfnamefont {R.}~\bibnamefont
  {Peng}}, \bibinfo {author} {\bibfnamefont {X.~P.}\ \bibnamefont {Shen}},
  \bibinfo {author} {\bibfnamefont {X.}~\bibnamefont {Xie}}, \bibinfo {author}
  {\bibfnamefont {H.~C.}\ \bibnamefont {Xu}}, \bibinfo {author} {\bibfnamefont
  {S.~Y.}\ \bibnamefont {Tan}}, \bibinfo {author} {\bibfnamefont
  {M.}~\bibnamefont {Xia}}, \bibinfo {author} {\bibfnamefont {T.}~\bibnamefont
  {Zhang}}, \bibinfo {author} {\bibfnamefont {H.~Y.}\ \bibnamefont {Cao}},
  \bibinfo {author} {\bibfnamefont {X.~G.}\ \bibnamefont {Gong}}, \bibinfo
  {author} {\bibfnamefont {J.~P.}\ \bibnamefont {Hu}}, \bibinfo {author}
  {\bibfnamefont {B.~P.}\ \bibnamefont {Xie}}, \ and\ \bibinfo {author}
  {\bibfnamefont {D.~L.}\ \bibnamefont {Feng}},\ }\href {\doibase
  10.1103/PhysRevLett.112.107001} {\bibfield  {journal} {\bibinfo  {journal}
  {Phys. Rev. Lett.}\ }\textbf {\bibinfo {volume} {112}},\ \bibinfo {pages}
  {107001} (\bibinfo {year} {2014}{\natexlab{a}})}\BibitemShut {NoStop}%
\bibitem [{\citenamefont {Li}\ \emph {et~al.}(2016)\citenamefont {Li},
  \citenamefont {Zhang}, \citenamefont {Tang}, \citenamefont {Liu},
  \citenamefont {Shi}, \citenamefont {Nie}, \citenamefont {Zhou}, \citenamefont
  {Li}, \citenamefont {Zhang}, \citenamefont {Song}, \citenamefont {He},
  \citenamefont {Ji}, \citenamefont {Zhang}, \citenamefont {Gu}, \citenamefont
  {Wang}, \citenamefont {Ma},\ and\ \citenamefont
  {Xue}}]{LiFangsen2015FeSe-structure}%
  \BibitemOpen
  \bibfield  {author} {\bibinfo {author} {\bibfnamefont {F.}~\bibnamefont
  {Li}}, \bibinfo {author} {\bibfnamefont {Q.}~\bibnamefont {Zhang}}, \bibinfo
  {author} {\bibfnamefont {C.}~\bibnamefont {Tang}}, \bibinfo {author}
  {\bibfnamefont {C.}~\bibnamefont {Liu}}, \bibinfo {author} {\bibfnamefont
  {J.}~\bibnamefont {Shi}}, \bibinfo {author} {\bibfnamefont {C.}~\bibnamefont
  {Nie}}, \bibinfo {author} {\bibfnamefont {G.}~\bibnamefont {Zhou}}, \bibinfo
  {author} {\bibfnamefont {Z.}~\bibnamefont {Li}}, \bibinfo {author}
  {\bibfnamefont {W.}~\bibnamefont {Zhang}}, \bibinfo {author} {\bibfnamefont
  {C.-L.}\ \bibnamefont {Song}}, \bibinfo {author} {\bibfnamefont
  {K.}~\bibnamefont {He}}, \bibinfo {author} {\bibfnamefont {S.}~\bibnamefont
  {Ji}}, \bibinfo {author} {\bibfnamefont {S.}~\bibnamefont {Zhang}}, \bibinfo
  {author} {\bibfnamefont {L.}~\bibnamefont {Gu}}, \bibinfo {author}
  {\bibfnamefont {L.}~\bibnamefont {Wang}}, \bibinfo {author} {\bibfnamefont
  {X.-C.}\ \bibnamefont {Ma}}, \ and\ \bibinfo {author} {\bibfnamefont {Q.-K.}\
  \bibnamefont {Xue}},\ }\href {http://stacks.iop.org/2053-1583/3/i=2/a=024002}
  {\bibfield  {journal} {\bibinfo  {journal} {2D Materials}\ }\textbf {\bibinfo
  {volume} {3}},\ \bibinfo {pages} {024002} (\bibinfo {year}
  {2016})}\BibitemShut {NoStop}%
\bibitem [{\citenamefont {Zou}\ \emph {et~al.}(2016)\citenamefont {Zou},
  \citenamefont {Mandal}, \citenamefont {Albright}, \citenamefont {Peng},
  \citenamefont {Pu}, \citenamefont {Kumah}, \citenamefont {Lau}, \citenamefont
  {Simon}, \citenamefont {Dagdeviren}, \citenamefont {He}, \citenamefont
  {Bo\ifmmode \check{z}\else \v{z}\fi{}ovi\ifmmode~\acute{c}\else \'{c}\fi{}},
  \citenamefont {Schwarz}, \citenamefont {Altman}, \citenamefont {Feng},
  \citenamefont {Walker}, \citenamefont {Ismail-Beigi},\ and\ \citenamefont
  {Ahn}}]{ZouPRB2016TiO2doublelayer}%
  \BibitemOpen
  \bibfield  {author} {\bibinfo {author} {\bibfnamefont {K.}~\bibnamefont
  {Zou}}, \bibinfo {author} {\bibfnamefont {S.}~\bibnamefont {Mandal}},
  \bibinfo {author} {\bibfnamefont {S.~D.}\ \bibnamefont {Albright}}, \bibinfo
  {author} {\bibfnamefont {R.}~\bibnamefont {Peng}}, \bibinfo {author}
  {\bibfnamefont {Y.}~\bibnamefont {Pu}}, \bibinfo {author} {\bibfnamefont
  {D.}~\bibnamefont {Kumah}}, \bibinfo {author} {\bibfnamefont
  {C.}~\bibnamefont {Lau}}, \bibinfo {author} {\bibfnamefont {G.~H.}\
  \bibnamefont {Simon}}, \bibinfo {author} {\bibfnamefont {O.~E.}\ \bibnamefont
  {Dagdeviren}}, \bibinfo {author} {\bibfnamefont {X.}~\bibnamefont {He}},
  \bibinfo {author} {\bibfnamefont {I.}~\bibnamefont {Bo\ifmmode \check{z}\else
  \v{z}\fi{}ovi\ifmmode~\acute{c}\else \'{c}\fi{}}}, \bibinfo {author}
  {\bibfnamefont {U.~D.}\ \bibnamefont {Schwarz}}, \bibinfo {author}
  {\bibfnamefont {E.~I.}\ \bibnamefont {Altman}}, \bibinfo {author}
  {\bibfnamefont {D.}~\bibnamefont {Feng}}, \bibinfo {author} {\bibfnamefont
  {F.~J.}\ \bibnamefont {Walker}}, \bibinfo {author} {\bibfnamefont
  {S.}~\bibnamefont {Ismail-Beigi}}, \ and\ \bibinfo {author} {\bibfnamefont
  {C.~H.}\ \bibnamefont {Ahn}},\ }\href {\doibase 10.1103/PhysRevB.93.180506}
  {\bibfield  {journal} {\bibinfo  {journal} {Phys. Rev. B}\ }\textbf {\bibinfo
  {volume} {93}},\ \bibinfo {pages} {180506} (\bibinfo {year}
  {2016})}\BibitemShut {NoStop}%
\bibitem [{\citenamefont {Peng}\ \emph
  {et~al.}(2014{\natexlab{b}})\citenamefont {Peng}, \citenamefont {Xu},
  \citenamefont {Tan}, \citenamefont {Cao}, \citenamefont {Xia}, \citenamefont
  {Shen}, \citenamefont {Huang}, \citenamefont {Wen}, \citenamefont {Song},
  \citenamefont {Zhang} \emph {et~al.}}]{PengRui2014NCFeSe}%
  \BibitemOpen
  \bibfield  {author} {\bibinfo {author} {\bibfnamefont {R.}~\bibnamefont
  {Peng}}, \bibinfo {author} {\bibfnamefont {H.}~\bibnamefont {Xu}}, \bibinfo
  {author} {\bibfnamefont {S.}~\bibnamefont {Tan}}, \bibinfo {author}
  {\bibfnamefont {H.}~\bibnamefont {Cao}}, \bibinfo {author} {\bibfnamefont
  {M.}~\bibnamefont {Xia}}, \bibinfo {author} {\bibfnamefont {X.}~\bibnamefont
  {Shen}}, \bibinfo {author} {\bibfnamefont {Z.}~\bibnamefont {Huang}},
  \bibinfo {author} {\bibfnamefont {C.}~\bibnamefont {Wen}}, \bibinfo {author}
  {\bibfnamefont {Q.}~\bibnamefont {Song}}, \bibinfo {author} {\bibfnamefont
  {T.}~\bibnamefont {Zhang}},  \emph {et~al.},\ }\href
  {http://www.nature.com/ncomms/2014/140926/ncomms6044/full/ncomms6044.html}
  {\bibfield  {journal} {\bibinfo  {journal} {Nature communications}\ }\textbf
  {\bibinfo {volume} {5}},\ \bibinfo {pages} {5044} (\bibinfo {year}
  {2014}{\natexlab{b}})}\BibitemShut {NoStop}%
\bibitem [{\citenamefont {Zhang}\ \emph
  {et~al.}(2016{\natexlab{a}})\citenamefont {Zhang}, \citenamefont {Peng},
  \citenamefont {Qian}, \citenamefont {Richard}, \citenamefont {Shi},
  \citenamefont {Ma}, \citenamefont {Fu}, \citenamefont {Guo}, \citenamefont
  {Han}, \citenamefont {Wang}, \citenamefont {Wang}, \citenamefont {Xue},
  \citenamefont {Hu}, \citenamefont {Sun},\ and\ \citenamefont
  {Ding}}]{Zhang2015FeSeSTO110}%
  \BibitemOpen
  \bibfield  {author} {\bibinfo {author} {\bibfnamefont {P.}~\bibnamefont
  {Zhang}}, \bibinfo {author} {\bibfnamefont {X.-L.}\ \bibnamefont {Peng}},
  \bibinfo {author} {\bibfnamefont {T.}~\bibnamefont {Qian}}, \bibinfo {author}
  {\bibfnamefont {P.}~\bibnamefont {Richard}}, \bibinfo {author} {\bibfnamefont
  {X.}~\bibnamefont {Shi}}, \bibinfo {author} {\bibfnamefont {J.-Z.}\
  \bibnamefont {Ma}}, \bibinfo {author} {\bibfnamefont {B.~B.}\ \bibnamefont
  {Fu}}, \bibinfo {author} {\bibfnamefont {Y.-L.}\ \bibnamefont {Guo}},
  \bibinfo {author} {\bibfnamefont {Z.~Q.}\ \bibnamefont {Han}}, \bibinfo
  {author} {\bibfnamefont {S.~C.}\ \bibnamefont {Wang}}, \bibinfo {author}
  {\bibfnamefont {L.~L.}\ \bibnamefont {Wang}}, \bibinfo {author}
  {\bibfnamefont {Q.-K.}\ \bibnamefont {Xue}}, \bibinfo {author} {\bibfnamefont
  {J.~P.}\ \bibnamefont {Hu}}, \bibinfo {author} {\bibfnamefont {Y.-J.}\
  \bibnamefont {Sun}}, \ and\ \bibinfo {author} {\bibfnamefont
  {H.}~\bibnamefont {Ding}},\ }\href {\doibase 10.1103/PhysRevB.94.104510}
  {\bibfield  {journal} {\bibinfo  {journal} {Phys. Rev. B}\ }\textbf {\bibinfo
  {volume} {94}},\ \bibinfo {pages} {104510} (\bibinfo {year}
  {2016}{\natexlab{a}})}\BibitemShut {NoStop}%
\bibitem [{\citenamefont {Zhou}\ \emph {et~al.}(2016)\citenamefont {Zhou},
  \citenamefont {Zhang}, \citenamefont {Liu}, \citenamefont {Tang},
  \citenamefont {Wang}, \citenamefont {Li}, \citenamefont {Song}, \citenamefont
  {Ji}, \citenamefont {He}, \citenamefont {Wang}, \citenamefont {Ma},\ and\
  \citenamefont {Xue}}]{ZhouGuanyu2015FeSeSTO110}%
  \BibitemOpen
  \bibfield  {author} {\bibinfo {author} {\bibfnamefont {G.}~\bibnamefont
  {Zhou}}, \bibinfo {author} {\bibfnamefont {D.}~\bibnamefont {Zhang}},
  \bibinfo {author} {\bibfnamefont {C.}~\bibnamefont {Liu}}, \bibinfo {author}
  {\bibfnamefont {C.}~\bibnamefont {Tang}}, \bibinfo {author} {\bibfnamefont
  {X.}~\bibnamefont {Wang}}, \bibinfo {author} {\bibfnamefont {Z.}~\bibnamefont
  {Li}}, \bibinfo {author} {\bibfnamefont {C.}~\bibnamefont {Song}}, \bibinfo
  {author} {\bibfnamefont {S.}~\bibnamefont {Ji}}, \bibinfo {author}
  {\bibfnamefont {K.}~\bibnamefont {He}}, \bibinfo {author} {\bibfnamefont
  {L.}~\bibnamefont {Wang}}, \bibinfo {author} {\bibfnamefont {X.}~\bibnamefont
  {Ma}}, \ and\ \bibinfo {author} {\bibfnamefont {Q.-K.}\ \bibnamefont {Xue}},\
  }\href
  {http://scitation.aip.org/content/aip/journal/apl/108/20/10.1063/1.4950964}
  {\bibfield  {journal} {\bibinfo  {journal} {Applied Physics Letters}\
  }\textbf {\bibinfo {volume} {108}},\ \bibinfo {eid} {202603} (\bibinfo {year}
  {2016})}\BibitemShut {NoStop}%
\bibitem [{\citenamefont {Ding}\ \emph {et~al.}(2016)\citenamefont {Ding},
  \citenamefont {Lv}, \citenamefont {Zhao}, \citenamefont {Wang}, \citenamefont
  {Wang}, \citenamefont {Song}, \citenamefont {Chen}, \citenamefont {Ma},\ and\
  \citenamefont {Xue}}]{DingHao2016FeSeTiO2}%
  \BibitemOpen
  \bibfield  {author} {\bibinfo {author} {\bibfnamefont {H.}~\bibnamefont
  {Ding}}, \bibinfo {author} {\bibfnamefont {Y.-F.}\ \bibnamefont {Lv}},
  \bibinfo {author} {\bibfnamefont {K.}~\bibnamefont {Zhao}}, \bibinfo {author}
  {\bibfnamefont {W.-L.}\ \bibnamefont {Wang}}, \bibinfo {author}
  {\bibfnamefont {L.}~\bibnamefont {Wang}}, \bibinfo {author} {\bibfnamefont
  {C.-L.}\ \bibnamefont {Song}}, \bibinfo {author} {\bibfnamefont
  {X.}~\bibnamefont {Chen}}, \bibinfo {author} {\bibfnamefont {X.-C.}\
  \bibnamefont {Ma}}, \ and\ \bibinfo {author} {\bibfnamefont {Q.-K.}\
  \bibnamefont {Xue}},\ }\href {\doibase 10.1103/PhysRevLett.117.067001}
  {\bibfield  {journal} {\bibinfo  {journal} {Phys. Rev. Lett.}\ }\textbf
  {\bibinfo {volume} {117}},\ \bibinfo {pages} {067001} (\bibinfo {year}
  {2016})}\BibitemShut {NoStop}%
\bibitem [{\citenamefont {Rebec}\ \emph {et~al.}(2017)\citenamefont {Rebec},
  \citenamefont {Jia}, \citenamefont {Zhang}, \citenamefont {Hashimoto},
  \citenamefont {Lu}, \citenamefont {Moore},\ and\ \citenamefont
  {Shen}}]{Rebec2017prlTiO2}%
  \BibitemOpen
  \bibfield  {author} {\bibinfo {author} {\bibfnamefont {S.~N.}\ \bibnamefont
  {Rebec}}, \bibinfo {author} {\bibfnamefont {T.}~\bibnamefont {Jia}}, \bibinfo
  {author} {\bibfnamefont {C.}~\bibnamefont {Zhang}}, \bibinfo {author}
  {\bibfnamefont {M.}~\bibnamefont {Hashimoto}}, \bibinfo {author}
  {\bibfnamefont {D.-H.}\ \bibnamefont {Lu}}, \bibinfo {author} {\bibfnamefont
  {R.~G.}\ \bibnamefont {Moore}}, \ and\ \bibinfo {author} {\bibfnamefont
  {Z.-X.}\ \bibnamefont {Shen}},\ }\href {\doibase
  10.1103/PhysRevLett.118.067002} {\bibfield  {journal} {\bibinfo  {journal}
  {Phys. Rev. Lett.}\ }\textbf {\bibinfo {volume} {118}},\ \bibinfo {pages}
  {067002} (\bibinfo {year} {2017})}\BibitemShut {NoStop}%
\bibitem [{\citenamefont {Tang}\ \emph {et~al.}(2016)\citenamefont {Tang},
  \citenamefont {Liu}, \citenamefont {Zhou}, \citenamefont {Li}, \citenamefont
  {Ding}, \citenamefont {Li}, \citenamefont {Zhang}, \citenamefont {Li},
  \citenamefont {Song}, \citenamefont {Ji}, \citenamefont {He}, \citenamefont
  {Wang}, \citenamefont {Ma},\ and\ \citenamefont {Xue}}]{TangChenjia2016FeSe}%
  \BibitemOpen
  \bibfield  {author} {\bibinfo {author} {\bibfnamefont {C.}~\bibnamefont
  {Tang}}, \bibinfo {author} {\bibfnamefont {C.}~\bibnamefont {Liu}}, \bibinfo
  {author} {\bibfnamefont {G.}~\bibnamefont {Zhou}}, \bibinfo {author}
  {\bibfnamefont {F.}~\bibnamefont {Li}}, \bibinfo {author} {\bibfnamefont
  {H.}~\bibnamefont {Ding}}, \bibinfo {author} {\bibfnamefont {Z.}~\bibnamefont
  {Li}}, \bibinfo {author} {\bibfnamefont {D.}~\bibnamefont {Zhang}}, \bibinfo
  {author} {\bibfnamefont {Z.}~\bibnamefont {Li}}, \bibinfo {author}
  {\bibfnamefont {C.}~\bibnamefont {Song}}, \bibinfo {author} {\bibfnamefont
  {S.}~\bibnamefont {Ji}}, \bibinfo {author} {\bibfnamefont {K.}~\bibnamefont
  {He}}, \bibinfo {author} {\bibfnamefont {L.}~\bibnamefont {Wang}}, \bibinfo
  {author} {\bibfnamefont {X.}~\bibnamefont {Ma}}, \ and\ \bibinfo {author}
  {\bibfnamefont {Q.-K.}\ \bibnamefont {Xue}},\ }\href {\doibase
  10.1103/PhysRevB.93.020507} {\bibfield  {journal} {\bibinfo  {journal} {Phys.
  Rev. B}\ }\textbf {\bibinfo {volume} {93}},\ \bibinfo {pages} {020507}
  (\bibinfo {year} {2016})}\BibitemShut {NoStop}%
\bibitem [{\citenamefont {Coh}\ \emph {et~al.}(2015)\citenamefont {Coh},
  \citenamefont {Cohen},\ and\ \citenamefont {Louie}}]{SinisaCoh2015FeSeECP}%
  \BibitemOpen
  \bibfield  {author} {\bibinfo {author} {\bibfnamefont {S.}~\bibnamefont
  {Coh}}, \bibinfo {author} {\bibfnamefont {M.~L.}\ \bibnamefont {Cohen}}, \
  and\ \bibinfo {author} {\bibfnamefont {S.~G.}\ \bibnamefont {Louie}},\ }\href
  {\doibase 10.1088/1367-2630/17/7/073027} {\bibfield  {journal} {\bibinfo
  {journal} {New Journal of Physics}\ }\textbf {\bibinfo {volume} {17}},\
  \bibinfo {pages} {073027} (\bibinfo {year} {2015})}\BibitemShut {NoStop}%
\bibitem [{\citenamefont {Wang}\ \emph
  {et~al.}(2016{\natexlab{b}})\citenamefont {Wang}, \citenamefont {Linscheid},
  \citenamefont {Berlijn},\ and\ \citenamefont {Johnston}}]{WangYan2016}%
  \BibitemOpen
  \bibfield  {author} {\bibinfo {author} {\bibfnamefont {Y.}~\bibnamefont
  {Wang}}, \bibinfo {author} {\bibfnamefont {A.}~\bibnamefont {Linscheid}},
  \bibinfo {author} {\bibfnamefont {T.}~\bibnamefont {Berlijn}}, \ and\
  \bibinfo {author} {\bibfnamefont {S.}~\bibnamefont {Johnston}},\ }\href
  {\doibase 10.1103/PhysRevB.93.134513} {\bibfield  {journal} {\bibinfo
  {journal} {Phys. Rev. B}\ }\textbf {\bibinfo {volume} {93}},\ \bibinfo
  {pages} {134513} (\bibinfo {year} {2016}{\natexlab{b}})}\BibitemShut
  {NoStop}%
\bibitem [{\citenamefont {Li}\ \emph {et~al.}(2014)\citenamefont {Li},
  \citenamefont {Xing}, \citenamefont {Huang},\ and\ \citenamefont
  {Xing}}]{DYXing2014FeSeSTOEPC}%
  \BibitemOpen
  \bibfield  {author} {\bibinfo {author} {\bibfnamefont {B.}~\bibnamefont
  {Li}}, \bibinfo {author} {\bibfnamefont {Z.~W.}\ \bibnamefont {Xing}},
  \bibinfo {author} {\bibfnamefont {G.~Q.}\ \bibnamefont {Huang}}, \ and\
  \bibinfo {author} {\bibfnamefont {D.~Y.}\ \bibnamefont {Xing}},\ }\href
  {\doibase 10.1063/1.4876750} {\bibfield  {journal} {\bibinfo  {journal}
  {Journal of Applied Physics}\ }\textbf {\bibinfo {volume} {115}},\ \bibinfo
  {pages} {193907} (\bibinfo {year} {2014})},\ \Eprint
  {http://arxiv.org/abs/http://dx.doi.org/10.1063/1.4876750}
  {http://dx.doi.org/10.1063/1.4876750} \BibitemShut {NoStop}%
\bibitem [{\citenamefont {Rademaker}\ \emph {et~al.}(2016)\citenamefont
  {Rademaker}, \citenamefont {Wang}, \citenamefont {Berlijn},\ and\
  \citenamefont {Johnston}}]{Johnstone2016PRBForwardscattering}%
  \BibitemOpen
  \bibfield  {author} {\bibinfo {author} {\bibfnamefont {L.}~\bibnamefont
  {Rademaker}}, \bibinfo {author} {\bibfnamefont {Y.}~\bibnamefont {Wang}},
  \bibinfo {author} {\bibfnamefont {T.}~\bibnamefont {Berlijn}}, \ and\
  \bibinfo {author} {\bibfnamefont {S.}~\bibnamefont {Johnston}},\ }\href
  {http://stacks.iop.org/1367-2630/18/i=2/a=022001} {\bibfield  {journal}
  {\bibinfo  {journal} {New Journal of Physics}\ }\textbf {\bibinfo {volume}
  {18}},\ \bibinfo {pages} {022001} (\bibinfo {year} {2016})}\BibitemShut
  {NoStop}%
\bibitem [{\citenamefont {Zhang}\ \emph
  {et~al.}(2016{\natexlab{b}})\citenamefont {Zhang}, \citenamefont {Guan},
  \citenamefont {Jia}, \citenamefont {Liu}, \citenamefont {Wang}, \citenamefont
  {Li}, \citenamefont {Wang}, \citenamefont {Ma}, \citenamefont {Xue},
  \citenamefont {Zhang}, \citenamefont {Plummer}, \citenamefont {Zhu},\ and\
  \citenamefont {Guo}}]{ShuyuanZhang2016}%
  \BibitemOpen
  \bibfield  {author} {\bibinfo {author} {\bibfnamefont {S.}~\bibnamefont
  {Zhang}}, \bibinfo {author} {\bibfnamefont {J.}~\bibnamefont {Guan}},
  \bibinfo {author} {\bibfnamefont {X.}~\bibnamefont {Jia}}, \bibinfo {author}
  {\bibfnamefont {B.}~\bibnamefont {Liu}}, \bibinfo {author} {\bibfnamefont
  {W.}~\bibnamefont {Wang}}, \bibinfo {author} {\bibfnamefont {F.}~\bibnamefont
  {Li}}, \bibinfo {author} {\bibfnamefont {L.}~\bibnamefont {Wang}}, \bibinfo
  {author} {\bibfnamefont {X.}~\bibnamefont {Ma}}, \bibinfo {author}
  {\bibfnamefont {Q.}~\bibnamefont {Xue}}, \bibinfo {author} {\bibfnamefont
  {J.}~\bibnamefont {Zhang}}, \bibinfo {author} {\bibfnamefont {E.~W.}\
  \bibnamefont {Plummer}}, \bibinfo {author} {\bibfnamefont {X.}~\bibnamefont
  {Zhu}}, \ and\ \bibinfo {author} {\bibfnamefont {J.}~\bibnamefont {Guo}},\
  }\href {\doibase 10.1103/PhysRevB.94.081116} {\bibfield  {journal} {\bibinfo
  {journal} {Phys. Rev. B}\ }\textbf {\bibinfo {volume} {94}},\ \bibinfo
  {pages} {081116} (\bibinfo {year} {2016}{\natexlab{b}})}\BibitemShut
  {NoStop}%
\bibitem [{\citenamefont {Wang}\ \emph
  {et~al.}(2016{\natexlab{c}})\citenamefont {Wang}, \citenamefont {Shen},
  \citenamefont {Pan}, \citenamefont {Hao}, \citenamefont {Ma}, \citenamefont
  {Zhou}, \citenamefont {Steffens}, \citenamefont {Schmalzl}, \citenamefont
  {Forrest}, \citenamefont {Abdel-Hafiez} \emph
  {et~al.}}]{wang2015SpinFluctuation}%
  \BibitemOpen
  \bibfield  {author} {\bibinfo {author} {\bibfnamefont {Q.}~\bibnamefont
  {Wang}}, \bibinfo {author} {\bibfnamefont {Y.}~\bibnamefont {Shen}}, \bibinfo
  {author} {\bibfnamefont {B.}~\bibnamefont {Pan}}, \bibinfo {author}
  {\bibfnamefont {Y.}~\bibnamefont {Hao}}, \bibinfo {author} {\bibfnamefont
  {M.}~\bibnamefont {Ma}}, \bibinfo {author} {\bibfnamefont {F.}~\bibnamefont
  {Zhou}}, \bibinfo {author} {\bibfnamefont {P.}~\bibnamefont {Steffens}},
  \bibinfo {author} {\bibfnamefont {K.}~\bibnamefont {Schmalzl}}, \bibinfo
  {author} {\bibfnamefont {T.}~\bibnamefont {Forrest}}, \bibinfo {author}
  {\bibfnamefont {M.}~\bibnamefont {Abdel-Hafiez}},  \emph {et~al.},\ }\href
  {\doibase doi:10.1038/nmat4492} {\bibfield  {journal} {\bibinfo  {journal}
  {Nature materials}\ }\textbf {\bibinfo {volume} {15}},\ \bibinfo {pages}
  {159} (\bibinfo {year} {2016}{\natexlab{c}})}\BibitemShut {NoStop}%
\bibitem [{\citenamefont {Wang}\ \emph
  {et~al.}(2016{\natexlab{d}})\citenamefont {Wang}, \citenamefont {Shen},
  \citenamefont {Pan}, \citenamefont {Zhang}, \citenamefont {Ikeuchi},
  \citenamefont {Iida}, \citenamefont {Christianson}, \citenamefont {Walker},
  \citenamefont {Adroja}, \citenamefont {Abdel-Hafiez} \emph
  {et~al.}}]{wang2016SpinFluctuationNC}%
  \BibitemOpen
  \bibfield  {author} {\bibinfo {author} {\bibfnamefont {Q.}~\bibnamefont
  {Wang}}, \bibinfo {author} {\bibfnamefont {Y.}~\bibnamefont {Shen}}, \bibinfo
  {author} {\bibfnamefont {B.}~\bibnamefont {Pan}}, \bibinfo {author}
  {\bibfnamefont {X.}~\bibnamefont {Zhang}}, \bibinfo {author} {\bibfnamefont
  {K.}~\bibnamefont {Ikeuchi}}, \bibinfo {author} {\bibfnamefont
  {K.}~\bibnamefont {Iida}}, \bibinfo {author} {\bibfnamefont {A.}~\bibnamefont
  {Christianson}}, \bibinfo {author} {\bibfnamefont {H.}~\bibnamefont
  {Walker}}, \bibinfo {author} {\bibfnamefont {D.}~\bibnamefont {Adroja}},
  \bibinfo {author} {\bibfnamefont {M.}~\bibnamefont {Abdel-Hafiez}},  \emph
  {et~al.},\ }\href {\doibase 10.1038/ncomms12182} {\bibfield  {journal}
  {\bibinfo  {journal} {Nature communications}\ }\textbf {\bibinfo {volume}
  {7}},\ \bibinfo {pages} {12182} (\bibinfo {year}
  {2016}{\natexlab{d}})}\BibitemShut {NoStop}%
\bibitem [{\citenamefont {Shukla}\ \emph {et~al.}(2003)\citenamefont {Shukla},
  \citenamefont {Calandra}, \citenamefont {d'Astuto}, \citenamefont {Lazzeri},
  \citenamefont {Mauri}, \citenamefont {Bellin}, \citenamefont {Krisch},
  \citenamefont {Karpinski}, \citenamefont {Kazakov}, \citenamefont {Jun},
  \citenamefont {Daghero},\ and\ \citenamefont
  {Parlinski}}]{Shukla2003MgB2LinewidthPRL}%
  \BibitemOpen
  \bibfield  {author} {\bibinfo {author} {\bibfnamefont {A.}~\bibnamefont
  {Shukla}}, \bibinfo {author} {\bibfnamefont {M.}~\bibnamefont {Calandra}},
  \bibinfo {author} {\bibfnamefont {M.}~\bibnamefont {d'Astuto}}, \bibinfo
  {author} {\bibfnamefont {M.}~\bibnamefont {Lazzeri}}, \bibinfo {author}
  {\bibfnamefont {F.}~\bibnamefont {Mauri}}, \bibinfo {author} {\bibfnamefont
  {C.}~\bibnamefont {Bellin}}, \bibinfo {author} {\bibfnamefont
  {M.}~\bibnamefont {Krisch}}, \bibinfo {author} {\bibfnamefont
  {J.}~\bibnamefont {Karpinski}}, \bibinfo {author} {\bibfnamefont {S.~M.}\
  \bibnamefont {Kazakov}}, \bibinfo {author} {\bibfnamefont {J.}~\bibnamefont
  {Jun}}, \bibinfo {author} {\bibfnamefont {D.}~\bibnamefont {Daghero}}, \ and\
  \bibinfo {author} {\bibfnamefont {K.}~\bibnamefont {Parlinski}},\ }\href
  {\doibase 10.1103/PhysRevLett.90.095506} {\bibfield  {journal} {\bibinfo
  {journal} {Phys. Rev. Lett.}\ }\textbf {\bibinfo {volume} {90}},\ \bibinfo
  {pages} {095506} (\bibinfo {year} {2003})}\BibitemShut {NoStop}%
\bibitem [{\citenamefont {Gnezdilov}\ \emph {et~al.}(2013)\citenamefont
  {Gnezdilov}, \citenamefont {Pashkevich}, \citenamefont {Lemmens},
  \citenamefont {Wulferding}, \citenamefont {Shevtsova}, \citenamefont {Gusev},
  \citenamefont {Chareev},\ and\ \citenamefont
  {Vasiliev}}]{Gnezdilov2013FeSePhonon}%
  \BibitemOpen
  \bibfield  {author} {\bibinfo {author} {\bibfnamefont {V.}~\bibnamefont
  {Gnezdilov}}, \bibinfo {author} {\bibfnamefont {Y.~G.}\ \bibnamefont
  {Pashkevich}}, \bibinfo {author} {\bibfnamefont {P.}~\bibnamefont {Lemmens}},
  \bibinfo {author} {\bibfnamefont {D.}~\bibnamefont {Wulferding}}, \bibinfo
  {author} {\bibfnamefont {T.}~\bibnamefont {Shevtsova}}, \bibinfo {author}
  {\bibfnamefont {A.}~\bibnamefont {Gusev}}, \bibinfo {author} {\bibfnamefont
  {D.}~\bibnamefont {Chareev}}, \ and\ \bibinfo {author} {\bibfnamefont
  {A.}~\bibnamefont {Vasiliev}},\ }\href {\doibase 10.1103/PhysRevB.87.144508}
  {\bibfield  {journal} {\bibinfo  {journal} {Phys. Rev. B}\ }\textbf {\bibinfo
  {volume} {87}},\ \bibinfo {pages} {144508} (\bibinfo {year}
  {2013})}\BibitemShut {NoStop}%
\bibitem [{\citenamefont {Mittal}\ \emph {et~al.}(2009)\citenamefont {Mittal},
  \citenamefont {Pintschovius}, \citenamefont {Lamago}, \citenamefont {Heid},
  \citenamefont {Bohnen}, \citenamefont {Reznik}, \citenamefont {Chaplot},
  \citenamefont {Su}, \citenamefont {Kumar}, \citenamefont {Dhar},
  \citenamefont {Thamizhavel},\ and\ \citenamefont
  {Brueckel}}]{NeutronPRL2009CaFeAs}%
  \BibitemOpen
  \bibfield  {author} {\bibinfo {author} {\bibfnamefont {R.}~\bibnamefont
  {Mittal}}, \bibinfo {author} {\bibfnamefont {L.}~\bibnamefont
  {Pintschovius}}, \bibinfo {author} {\bibfnamefont {D.}~\bibnamefont
  {Lamago}}, \bibinfo {author} {\bibfnamefont {R.}~\bibnamefont {Heid}},
  \bibinfo {author} {\bibfnamefont {K.-P.}\ \bibnamefont {Bohnen}}, \bibinfo
  {author} {\bibfnamefont {D.}~\bibnamefont {Reznik}}, \bibinfo {author}
  {\bibfnamefont {S.~L.}\ \bibnamefont {Chaplot}}, \bibinfo {author}
  {\bibfnamefont {Y.}~\bibnamefont {Su}}, \bibinfo {author} {\bibfnamefont
  {N.}~\bibnamefont {Kumar}}, \bibinfo {author} {\bibfnamefont {S.~K.}\
  \bibnamefont {Dhar}}, \bibinfo {author} {\bibfnamefont {A.}~\bibnamefont
  {Thamizhavel}}, \ and\ \bibinfo {author} {\bibfnamefont {T.}~\bibnamefont
  {Brueckel}},\ }\href {\doibase 10.1103/PhysRevLett.102.217001} {\bibfield
  {journal} {\bibinfo  {journal} {Phys. Rev. Lett.}\ }\textbf {\bibinfo
  {volume} {102}},\ \bibinfo {pages} {217001} (\bibinfo {year}
  {2009})}\BibitemShut {NoStop}%
\bibitem [{\citenamefont {Ksenofontov}\ \emph {et~al.}(2010)\citenamefont
  {Ksenofontov}, \citenamefont {Wortmann}, \citenamefont {Chumakov},
  \citenamefont {Gasi}, \citenamefont {Medvedev}, \citenamefont {McQueen},
  \citenamefont {Cava},\ and\ \citenamefont {Felser}}]{FeSeDebye2010prb}%
  \BibitemOpen
  \bibfield  {author} {\bibinfo {author} {\bibfnamefont {V.}~\bibnamefont
  {Ksenofontov}}, \bibinfo {author} {\bibfnamefont {G.}~\bibnamefont
  {Wortmann}}, \bibinfo {author} {\bibfnamefont {A.~I.}\ \bibnamefont
  {Chumakov}}, \bibinfo {author} {\bibfnamefont {T.}~\bibnamefont {Gasi}},
  \bibinfo {author} {\bibfnamefont {S.}~\bibnamefont {Medvedev}}, \bibinfo
  {author} {\bibfnamefont {T.~M.}\ \bibnamefont {McQueen}}, \bibinfo {author}
  {\bibfnamefont {R.~J.}\ \bibnamefont {Cava}}, \ and\ \bibinfo {author}
  {\bibfnamefont {C.}~\bibnamefont {Felser}},\ }\href {\doibase
  10.1103/PhysRevB.81.184510} {\bibfield  {journal} {\bibinfo  {journal} {Phys.
  Rev. B}\ }\textbf {\bibinfo {volume} {81}},\ \bibinfo {pages} {184510}
  (\bibinfo {year} {2010})}\BibitemShut {NoStop}%
\bibitem [{\citenamefont {Gang}\ \emph {et~al.}(2008)\citenamefont {Gang},
  \citenamefont {Xi-Yu}, \citenamefont {Lei}, \citenamefont {Lei},
  \citenamefont {Cong},\ and\ \citenamefont
  {Hai-Hu}}]{LaOFFeAsCPL2008specificheat}%
  \BibitemOpen
  \bibfield  {author} {\bibinfo {author} {\bibfnamefont {M.}~\bibnamefont
  {Gang}}, \bibinfo {author} {\bibfnamefont {Z.}~\bibnamefont {Xi-Yu}},
  \bibinfo {author} {\bibfnamefont {F.}~\bibnamefont {Lei}}, \bibinfo {author}
  {\bibfnamefont {S.}~\bibnamefont {Lei}}, \bibinfo {author} {\bibfnamefont
  {R.}~\bibnamefont {Cong}}, \ and\ \bibinfo {author} {\bibfnamefont
  {W.}~\bibnamefont {Hai-Hu}},\ }\href
  {http://stacks.iop.org/0256-307X/25/i=6/a=082} {\bibfield  {journal}
  {\bibinfo  {journal} {Chinese Physics Letters}\ }\textbf {\bibinfo {volume}
  {25}},\ \bibinfo {pages} {2221} (\bibinfo {year} {2008})}\BibitemShut
  {NoStop}%
\bibitem [{\citenamefont {Lin}\ \emph {et~al.}(2011)\citenamefont {Lin},
  \citenamefont {Hsieh}, \citenamefont {Chareev}, \citenamefont {Vasiliev},
  \citenamefont {Parsons},\ and\ \citenamefont {Yang}}]{FeSeSpecificheat}%
  \BibitemOpen
  \bibfield  {author} {\bibinfo {author} {\bibfnamefont {J.-Y.}\ \bibnamefont
  {Lin}}, \bibinfo {author} {\bibfnamefont {Y.~S.}\ \bibnamefont {Hsieh}},
  \bibinfo {author} {\bibfnamefont {D.~A.}\ \bibnamefont {Chareev}}, \bibinfo
  {author} {\bibfnamefont {A.~N.}\ \bibnamefont {Vasiliev}}, \bibinfo {author}
  {\bibfnamefont {Y.}~\bibnamefont {Parsons}}, \ and\ \bibinfo {author}
  {\bibfnamefont {H.~D.}\ \bibnamefont {Yang}},\ }\href {\doibase
  10.1103/PhysRevB.84.220507} {\bibfield  {journal} {\bibinfo  {journal} {Phys.
  Rev. B}\ }\textbf {\bibinfo {volume} {84}},\ \bibinfo {pages} {220507}
  (\bibinfo {year} {2011})}\BibitemShut {NoStop}%
\bibitem [{\citenamefont {Lawless}\ and\ \citenamefont
  {Morrow}(1977)}]{Lawless1977Debye}%
  \BibitemOpen
  \bibfield  {author} {\bibinfo {author} {\bibfnamefont {W.~N.}\ \bibnamefont
  {Lawless}}\ and\ \bibinfo {author} {\bibfnamefont {A.~J.}\ \bibnamefont
  {Morrow}},\ }\href {\doibase 10.1080/00150197708237810} {\bibfield  {journal}
  {\bibinfo  {journal} {Ferroelectrics}\ }\textbf {\bibinfo {volume} {15}},\
  \bibinfo {pages} {159} (\bibinfo {year} {1977})},\ \Eprint
  {http://arxiv.org/abs/http://dx.doi.org/10.1080/00150197708237810}
  {http://dx.doi.org/10.1080/00150197708237810} \BibitemShut {NoStop}%
\bibitem [{\citenamefont {Bell}\ and\ \citenamefont
  {Rupprecht}(1963)}]{ElasticConstantsSTO}%
  \BibitemOpen
  \bibfield  {author} {\bibinfo {author} {\bibfnamefont {R.~O.}\ \bibnamefont
  {Bell}}\ and\ \bibinfo {author} {\bibfnamefont {G.}~\bibnamefont
  {Rupprecht}},\ }\href {\doibase 10.1103/PhysRev.129.90} {\bibfield  {journal}
  {\bibinfo  {journal} {Phys. Rev.}\ }\textbf {\bibinfo {volume} {129}},\
  \bibinfo {pages} {90} (\bibinfo {year} {1963})}\BibitemShut {NoStop}%
\bibitem [{\citenamefont {Benedek}\ \emph {et~al.}(2013)\citenamefont
  {Benedek}, \citenamefont {Celli}, \citenamefont {Comsa}, \citenamefont
  {Doak}, \citenamefont {Frenken}, \citenamefont {Hinch}, \citenamefont
  {Hoinkes}, \citenamefont {Kern}, \citenamefont {Lahee}, \citenamefont
  {Lapujoulade} \emph {et~al.}}]{benedek2013heliumBook}%
  \BibitemOpen
  \bibfield  {author} {\bibinfo {author} {\bibfnamefont {G.}~\bibnamefont
  {Benedek}}, \bibinfo {author} {\bibfnamefont {V.}~\bibnamefont {Celli}},
  \bibinfo {author} {\bibfnamefont {G.}~\bibnamefont {Comsa}}, \bibinfo
  {author} {\bibfnamefont {R.}~\bibnamefont {Doak}}, \bibinfo {author}
  {\bibfnamefont {J.}~\bibnamefont {Frenken}}, \bibinfo {author} {\bibfnamefont
  {B.}~\bibnamefont {Hinch}}, \bibinfo {author} {\bibfnamefont
  {H.}~\bibnamefont {Hoinkes}}, \bibinfo {author} {\bibfnamefont
  {K.}~\bibnamefont {Kern}}, \bibinfo {author} {\bibfnamefont {A.}~\bibnamefont
  {Lahee}}, \bibinfo {author} {\bibfnamefont {J.}~\bibnamefont {Lapujoulade}},
  \emph {et~al.},\ }\href@noop {} {\emph {\bibinfo {title} {Helium atom
  scattering from surfaces}}},\ Vol.~\bibinfo {volume} {27}\ (\bibinfo
  {publisher} {Springer Science \& Business Media},\ \bibinfo {year}
  {2013})\BibitemShut {NoStop}%
\bibitem [{\citenamefont {Phelan}\ \emph {et~al.}(2009)\citenamefont {Phelan},
  \citenamefont {Millican}, \citenamefont {Thomas}, \citenamefont {Le\~ao},
  \citenamefont {Qiu},\ and\ \citenamefont {Paul}}]{Phelan2009FeSePhononDOS}%
  \BibitemOpen
  \bibfield  {author} {\bibinfo {author} {\bibfnamefont {D.}~\bibnamefont
  {Phelan}}, \bibinfo {author} {\bibfnamefont {J.~N.}\ \bibnamefont
  {Millican}}, \bibinfo {author} {\bibfnamefont {E.~L.}\ \bibnamefont
  {Thomas}}, \bibinfo {author} {\bibfnamefont {J.~B.}\ \bibnamefont {Le\~ao}},
  \bibinfo {author} {\bibfnamefont {Y.}~\bibnamefont {Qiu}}, \ and\ \bibinfo
  {author} {\bibfnamefont {R.}~\bibnamefont {Paul}},\ }\href {\doibase
  10.1103/PhysRevB.79.014519} {\bibfield  {journal} {\bibinfo  {journal} {Phys.
  Rev. B}\ }\textbf {\bibinfo {volume} {79}},\ \bibinfo {pages} {014519}
  (\bibinfo {year} {2009})}\BibitemShut {NoStop}%
\bibitem [{\citenamefont {Cao}\ \emph {et~al.}(2012)\citenamefont {Cao},
  \citenamefont {Wang}, \citenamefont {Liu}, \citenamefont {Guo},\ and\
  \citenamefont {Guo}}]{CaoYanweiSTO110}%
  \BibitemOpen
  \bibfield  {author} {\bibinfo {author} {\bibfnamefont {Y.}~\bibnamefont
  {Cao}}, \bibinfo {author} {\bibfnamefont {S.}~\bibnamefont {Wang}}, \bibinfo
  {author} {\bibfnamefont {S.}~\bibnamefont {Liu}}, \bibinfo {author}
  {\bibfnamefont {Q.}~\bibnamefont {Guo}}, \ and\ \bibinfo {author}
  {\bibfnamefont {J.}~\bibnamefont {Guo}},\ }\href {\doibase 10.1063/1.4737946}
  {\bibfield  {journal} {\bibinfo  {journal} {The Journal of Chemical Physics}\
  }\textbf {\bibinfo {volume} {137}},\ \bibinfo {pages} {044701} (\bibinfo
  {year} {2012})},\ \Eprint
  {http://arxiv.org/abs/http://dx.doi.org/10.1063/1.4737946}
  {http://dx.doi.org/10.1063/1.4737946} \BibitemShut {NoStop}%
\bibitem [{\citenamefont {Zhu}\ \emph {et~al.}(2015)\citenamefont {Zhu},
  \citenamefont {Cao}, \citenamefont {Zhang}, \citenamefont {Jia},
  \citenamefont {Guo}, \citenamefont {Yang}, \citenamefont {Zhu}, \citenamefont
  {Zhang}, \citenamefont {Plummer},\ and\ \citenamefont
  {Guo}}]{ZhuXuetao2015HREELS}%
  \BibitemOpen
  \bibfield  {author} {\bibinfo {author} {\bibfnamefont {X.}~\bibnamefont
  {Zhu}}, \bibinfo {author} {\bibfnamefont {Y.}~\bibnamefont {Cao}}, \bibinfo
  {author} {\bibfnamefont {S.}~\bibnamefont {Zhang}}, \bibinfo {author}
  {\bibfnamefont {X.}~\bibnamefont {Jia}}, \bibinfo {author} {\bibfnamefont
  {Q.}~\bibnamefont {Guo}}, \bibinfo {author} {\bibfnamefont {F.}~\bibnamefont
  {Yang}}, \bibinfo {author} {\bibfnamefont {L.}~\bibnamefont {Zhu}}, \bibinfo
  {author} {\bibfnamefont {J.}~\bibnamefont {Zhang}}, \bibinfo {author}
  {\bibfnamefont {E.}~\bibnamefont {Plummer}}, \ and\ \bibinfo {author}
  {\bibfnamefont {J.}~\bibnamefont {Guo}},\ }\href {\doibase 10.1063/1.4928215}
  {\bibfield  {journal} {\bibinfo  {journal} {Review of Scientific
  Instruments}\ }\textbf {\bibinfo {volume} {86}},\ \bibinfo {pages} {083902}
  (\bibinfo {year} {2015})}\BibitemShut {NoStop}%
\bibitem [{\citenamefont {Tabor}\ \emph {et~al.}(1971)\citenamefont {Tabor},
  \citenamefont {Wilson},\ and\ \citenamefont {Bastow}}]{Tabor1971LEEDDebye}%
  \BibitemOpen
  \bibfield  {author} {\bibinfo {author} {\bibfnamefont {D.}~\bibnamefont
  {Tabor}}, \bibinfo {author} {\bibfnamefont {J.}~\bibnamefont {Wilson}}, \
  and\ \bibinfo {author} {\bibfnamefont {T.}~\bibnamefont {Bastow}},\ }\href
  {\doibase http://dx.doi.org/10.1016/0039-6028(71)90009-4} {\bibfield
  {journal} {\bibinfo  {journal} {Surface Science}\ }\textbf {\bibinfo {volume}
  {26}},\ \bibinfo {pages} {471 } (\bibinfo {year} {1971})}\BibitemShut
  {NoStop}%
\bibitem [{\citenamefont {Jones}\ \emph {et~al.}(1966)\citenamefont {Jones},
  \citenamefont {McKinney},\ and\ \citenamefont {Webb}}]{Jones1966Debye}%
  \BibitemOpen
  \bibfield  {author} {\bibinfo {author} {\bibfnamefont {E.~R.}\ \bibnamefont
  {Jones}}, \bibinfo {author} {\bibfnamefont {J.~T.}\ \bibnamefont {McKinney}},
  \ and\ \bibinfo {author} {\bibfnamefont {M.~B.}\ \bibnamefont {Webb}},\
  }\href {\doibase 10.1103/PhysRev.151.476} {\bibfield  {journal} {\bibinfo
  {journal} {Phys. Rev.}\ }\textbf {\bibinfo {volume} {151}},\ \bibinfo {pages}
  {476} (\bibinfo {year} {1966})}\BibitemShut {NoStop}%
\bibitem [{\citenamefont {Reid}(1970)}]{reid1970debye}%
  \BibitemOpen
  \bibfield  {author} {\bibinfo {author} {\bibfnamefont {R.}~\bibnamefont
  {Reid}},\ }\href {\doibase 10.1002/pssa.19700020243} {\bibfield  {journal}
  {\bibinfo  {journal} {physica status solidi (a)}\ }\textbf {\bibinfo {volume}
  {2}},\ \bibinfo {pages} {K109} (\bibinfo {year} {1970})}\BibitemShut
  {NoStop}%
\bibitem [{\citenamefont {Debye}(1913)}]{Debye1913}%
  \BibitemOpen
  \bibfield  {author} {\bibinfo {author} {\bibfnamefont {P.}~\bibnamefont
  {Debye}},\ }\href {\doibase 10.1002/andp.19133480105} {\bibfield  {journal}
  {\bibinfo  {journal} {Annalen der Physik}\ }\textbf {\bibinfo {volume}
  {348}},\ \bibinfo {pages} {49} (\bibinfo {year} {1913})}\BibitemShut
  {NoStop}%
\bibitem [{\citenamefont {Waller}(1923)}]{Waller1923}%
  \BibitemOpen
  \bibfield  {author} {\bibinfo {author} {\bibfnamefont {I.}~\bibnamefont
  {Waller}},\ }\href {\doibase 10.1007/BF01328696} {\bibfield  {journal}
  {\bibinfo  {journal} {Zeitschrift f{\"u}r Physik}\ }\textbf {\bibinfo
  {volume} {17}},\ \bibinfo {pages} {398} (\bibinfo {year} {1923})}\BibitemShut
  {NoStop}%
\bibitem [{\citenamefont {James}(1952)}]{DebyeWaller1952james}%
  \BibitemOpen
  \bibfield  {author} {\bibinfo {author} {\bibfnamefont {R.}~\bibnamefont
  {James}},\ }\href@noop {} {\emph {\bibinfo {title} {The Optical Principles of
  the Diffraction of X-rays.}}}\ (\bibinfo  {publisher} {G. Bell and sons},\
  \bibinfo {year} {1952})\BibitemShut {NoStop}%
\bibitem [{\citenamefont {Ashcroft}\ and\ \citenamefont
  {Mermin}(1976)}]{ashcroft1976solid}%
  \BibitemOpen
  \bibfield  {author} {\bibinfo {author} {\bibfnamefont {N.~W.}\ \bibnamefont
  {Ashcroft}}\ and\ \bibinfo {author} {\bibfnamefont {N.~D.}\ \bibnamefont
  {Mermin}},\ }\href@noop {} {\emph {\bibinfo {title} {Solid state physics
  (holt, rinehart and winston, new york, 1976)}}}\ (\bibinfo {year} {1976})\
  p.\ \bibinfo {pages} {793}\BibitemShut {NoStop}%
\bibitem [{\citenamefont {Togo}\ and\ \citenamefont {Tanaka}(2015)}]{Togo2015}%
  \BibitemOpen
  \bibfield  {author} {\bibinfo {author} {\bibfnamefont {A.}~\bibnamefont
  {Togo}}\ and\ \bibinfo {author} {\bibfnamefont {I.}~\bibnamefont {Tanaka}},\
  }\href {\doibase https://doi.org/10.1016/j.scriptamat.2015.07.021} {\bibfield
   {journal} {\bibinfo  {journal} {Scripta Materialia}\ }\textbf {\bibinfo
  {volume} {108}},\ \bibinfo {pages} {1 } (\bibinfo {year} {2015})}\BibitemShut
  {NoStop}%
\bibitem [{\citenamefont {Bl\"ochl}(1994)}]{Bloechl1994}%
  \BibitemOpen
  \bibfield  {author} {\bibinfo {author} {\bibfnamefont {P.~E.}\ \bibnamefont
  {Bl\"ochl}},\ }\href {\doibase 10.1103/PhysRevB.50.17953} {\bibfield
  {journal} {\bibinfo  {journal} {Phys. Rev. B}\ }\textbf {\bibinfo {volume}
  {50}},\ \bibinfo {pages} {17953} (\bibinfo {year} {1994})}\BibitemShut
  {NoStop}%
\bibitem [{\citenamefont {Kresse}\ and\ \citenamefont
  {Furthm\"uller}(1996)}]{Kresse1996}%
  \BibitemOpen
  \bibfield  {author} {\bibinfo {author} {\bibfnamefont {G.}~\bibnamefont
  {Kresse}}\ and\ \bibinfo {author} {\bibfnamefont {J.}~\bibnamefont
  {Furthm\"uller}},\ }\href {\doibase 10.1103/PhysRevB.54.11169} {\bibfield
  {journal} {\bibinfo  {journal} {Phys. Rev. B}\ }\textbf {\bibinfo {volume}
  {54}},\ \bibinfo {pages} {11169} (\bibinfo {year} {1996})}\BibitemShut
  {NoStop}%
\bibitem [{\citenamefont {Kresse}\ and\ \citenamefont
  {Joubert}(1999)}]{Kresse1999}%
  \BibitemOpen
  \bibfield  {author} {\bibinfo {author} {\bibfnamefont {G.}~\bibnamefont
  {Kresse}}\ and\ \bibinfo {author} {\bibfnamefont {D.}~\bibnamefont
  {Joubert}},\ }\href {\doibase 10.1103/PhysRevB.59.1758} {\bibfield  {journal}
  {\bibinfo  {journal} {Phys. Rev. B}\ }\textbf {\bibinfo {volume} {59}},\
  \bibinfo {pages} {1758} (\bibinfo {year} {1999})}\BibitemShut {NoStop}%
\bibitem [{\citenamefont {Giannozzi}\ \emph {et~al.}(2009)\citenamefont
  {Giannozzi}, \citenamefont {Baroni}, \citenamefont {Bonini}, \citenamefont
  {Calandra}, \citenamefont {Car}, \citenamefont {Cavazzoni}, \citenamefont
  {Ceresoli}, \citenamefont {Chiarotti}, \citenamefont {Cococcioni},
  \citenamefont {Dabo}, \citenamefont {Corso}, \citenamefont {de~Gironcoli},
  \citenamefont {Fabris}, \citenamefont {Fratesi}, \citenamefont {Gebauer},
  \citenamefont {Gerstmann}, \citenamefont {Gougoussis}, \citenamefont
  {Kokalj}, \citenamefont {Lazzeri}, \citenamefont {Martin-Samos},
  \citenamefont {Marzari}, \citenamefont {Mauri}, \citenamefont {Mazzarello},
  \citenamefont {Paolini}, \citenamefont {Pasquarello}, \citenamefont
  {Paulatto}, \citenamefont {Sbraccia}, \citenamefont {Scandolo}, \citenamefont
  {Sclauzero}, \citenamefont {Seitsonen}, \citenamefont {Smogunov},
  \citenamefont {Umari},\ and\ \citenamefont {Wentzcovitch}}]{Giannozzi2009}%
  \BibitemOpen
  \bibfield  {author} {\bibinfo {author} {\bibfnamefont {P.}~\bibnamefont
  {Giannozzi}}, \bibinfo {author} {\bibfnamefont {S.}~\bibnamefont {Baroni}},
  \bibinfo {author} {\bibfnamefont {N.}~\bibnamefont {Bonini}}, \bibinfo
  {author} {\bibfnamefont {M.}~\bibnamefont {Calandra}}, \bibinfo {author}
  {\bibfnamefont {R.}~\bibnamefont {Car}}, \bibinfo {author} {\bibfnamefont
  {C.}~\bibnamefont {Cavazzoni}}, \bibinfo {author} {\bibfnamefont
  {D.}~\bibnamefont {Ceresoli}}, \bibinfo {author} {\bibfnamefont {G.~L.}\
  \bibnamefont {Chiarotti}}, \bibinfo {author} {\bibfnamefont {M.}~\bibnamefont
  {Cococcioni}}, \bibinfo {author} {\bibfnamefont {I.}~\bibnamefont {Dabo}},
  \bibinfo {author} {\bibfnamefont {A.~D.}\ \bibnamefont {Corso}}, \bibinfo
  {author} {\bibfnamefont {S.}~\bibnamefont {de~Gironcoli}}, \bibinfo {author}
  {\bibfnamefont {S.}~\bibnamefont {Fabris}}, \bibinfo {author} {\bibfnamefont
  {G.}~\bibnamefont {Fratesi}}, \bibinfo {author} {\bibfnamefont
  {R.}~\bibnamefont {Gebauer}}, \bibinfo {author} {\bibfnamefont
  {U.}~\bibnamefont {Gerstmann}}, \bibinfo {author} {\bibfnamefont
  {C.}~\bibnamefont {Gougoussis}}, \bibinfo {author} {\bibfnamefont
  {A.}~\bibnamefont {Kokalj}}, \bibinfo {author} {\bibfnamefont
  {M.}~\bibnamefont {Lazzeri}}, \bibinfo {author} {\bibfnamefont
  {L.}~\bibnamefont {Martin-Samos}}, \bibinfo {author} {\bibfnamefont
  {N.}~\bibnamefont {Marzari}}, \bibinfo {author} {\bibfnamefont
  {F.}~\bibnamefont {Mauri}}, \bibinfo {author} {\bibfnamefont
  {R.}~\bibnamefont {Mazzarello}}, \bibinfo {author} {\bibfnamefont
  {S.}~\bibnamefont {Paolini}}, \bibinfo {author} {\bibfnamefont
  {A.}~\bibnamefont {Pasquarello}}, \bibinfo {author} {\bibfnamefont
  {L.}~\bibnamefont {Paulatto}}, \bibinfo {author} {\bibfnamefont
  {C.}~\bibnamefont {Sbraccia}}, \bibinfo {author} {\bibfnamefont
  {S.}~\bibnamefont {Scandolo}}, \bibinfo {author} {\bibfnamefont
  {G.}~\bibnamefont {Sclauzero}}, \bibinfo {author} {\bibfnamefont {A.~P.}\
  \bibnamefont {Seitsonen}}, \bibinfo {author} {\bibfnamefont {A.}~\bibnamefont
  {Smogunov}}, \bibinfo {author} {\bibfnamefont {P.}~\bibnamefont {Umari}}, \
  and\ \bibinfo {author} {\bibfnamefont {R.~M.}\ \bibnamefont {Wentzcovitch}},\
  }\href {http://stacks.iop.org/0953-8984/21/i=39/a=395502} {\bibfield
  {journal} {\bibinfo  {journal} {Journal of Physics: Condensed Matter}\
  }\textbf {\bibinfo {volume} {21}},\ \bibinfo {pages} {395502} (\bibinfo
  {year} {2009})}\BibitemShut {NoStop}%
\bibitem [{\citenamefont {Garrity}\ \emph {et~al.}(2014)\citenamefont
  {Garrity}, \citenamefont {Bennett}, \citenamefont {Rabe},\ and\ \citenamefont
  {Vanderbilt}}]{Garrity2014}%
  \BibitemOpen
  \bibfield  {author} {\bibinfo {author} {\bibfnamefont {K.~F.}\ \bibnamefont
  {Garrity}}, \bibinfo {author} {\bibfnamefont {J.~W.}\ \bibnamefont
  {Bennett}}, \bibinfo {author} {\bibfnamefont {K.~M.}\ \bibnamefont {Rabe}}, \
  and\ \bibinfo {author} {\bibfnamefont {D.}~\bibnamefont {Vanderbilt}},\
  }\href {\doibase http://dx.doi.org/10.1016/j.commatsci.2013.08.053}
  {\bibfield  {journal} {\bibinfo  {journal} {Computational Materials Science}\
  }\textbf {\bibinfo {volume} {81}},\ \bibinfo {pages} {446 } (\bibinfo {year}
  {2014})}\BibitemShut {NoStop}%
\bibitem [{\citenamefont {Perdew}\ \emph {et~al.}(1996)\citenamefont {Perdew},
  \citenamefont {Burke},\ and\ \citenamefont {Ernzerhof}}]{Perdew1996}%
  \BibitemOpen
  \bibfield  {author} {\bibinfo {author} {\bibfnamefont {J.~P.}\ \bibnamefont
  {Perdew}}, \bibinfo {author} {\bibfnamefont {K.}~\bibnamefont {Burke}}, \
  and\ \bibinfo {author} {\bibfnamefont {M.}~\bibnamefont {Ernzerhof}},\ }\href
  {\doibase 10.1103/PhysRevLett.77.3865} {\bibfield  {journal} {\bibinfo
  {journal} {Phys. Rev. Lett.}\ }\textbf {\bibinfo {volume} {77}},\ \bibinfo
  {pages} {3865} (\bibinfo {year} {1996})}\BibitemShut {NoStop}%
\bibitem [{\citenamefont {Ibach}\ and\ \citenamefont
  {Mills}(1982)}]{ibach1982eels}%
  \BibitemOpen
  \bibfield  {author} {\bibinfo {author} {\bibfnamefont {H.}~\bibnamefont
  {Ibach}}\ and\ \bibinfo {author} {\bibfnamefont {D.~L.}\ \bibnamefont
  {Mills}},\ }\href@noop {} {\emph {\bibinfo {title} {Electron energy loss
  spectroscopy and surface vibrations}}}\ (\bibinfo  {publisher} {Academic
  press},\ \bibinfo {year} {1982})\BibitemShut {NoStop}%
\bibitem [{\citenamefont {Zhu}\ \emph {et~al.}(2011)\citenamefont {Zhu},
  \citenamefont {Santos}, \citenamefont {Sankar}, \citenamefont {Chikara},
  \citenamefont {Howard}, \citenamefont {Chou}, \citenamefont {Chamon},\ and\
  \citenamefont {El-Batanouny}}]{ZhuXuetao2012prlBi2Se3}%
  \BibitemOpen
  \bibfield  {author} {\bibinfo {author} {\bibfnamefont {X.}~\bibnamefont
  {Zhu}}, \bibinfo {author} {\bibfnamefont {L.}~\bibnamefont {Santos}},
  \bibinfo {author} {\bibfnamefont {R.}~\bibnamefont {Sankar}}, \bibinfo
  {author} {\bibfnamefont {S.}~\bibnamefont {Chikara}}, \bibinfo {author}
  {\bibfnamefont {C.~.}\ \bibnamefont {Howard}}, \bibinfo {author}
  {\bibfnamefont {F.~C.}\ \bibnamefont {Chou}}, \bibinfo {author}
  {\bibfnamefont {C.}~\bibnamefont {Chamon}}, \ and\ \bibinfo {author}
  {\bibfnamefont {M.}~\bibnamefont {El-Batanouny}},\ }\href {\doibase
  10.1103/PhysRevLett.107.186102} {\bibfield  {journal} {\bibinfo  {journal}
  {Phys. Rev. Lett.}\ }\textbf {\bibinfo {volume} {107}},\ \bibinfo {pages}
  {186102} (\bibinfo {year} {2011})}\BibitemShut {NoStop}%
\bibitem [{\citenamefont {McMillan}(1968)}]{McMillanEquation}%
  \BibitemOpen
  \bibfield  {author} {\bibinfo {author} {\bibfnamefont {W.~L.}\ \bibnamefont
  {McMillan}},\ }\href {\doibase 10.1103/PhysRev.167.331} {\bibfield  {journal}
  {\bibinfo  {journal} {Phys. Rev.}\ }\textbf {\bibinfo {volume} {167}},\
  \bibinfo {pages} {331} (\bibinfo {year} {1968})}\BibitemShut {NoStop}%
\bibitem [{\citenamefont {Boeri}\ \emph {et~al.}(2008)\citenamefont {Boeri},
  \citenamefont {Dolgov},\ and\ \citenamefont {Golubov}}]{Boeri2008ionEPC}%
  \BibitemOpen
  \bibfield  {author} {\bibinfo {author} {\bibfnamefont {L.}~\bibnamefont
  {Boeri}}, \bibinfo {author} {\bibfnamefont {O.~V.}\ \bibnamefont {Dolgov}}, \
  and\ \bibinfo {author} {\bibfnamefont {A.~A.}\ \bibnamefont {Golubov}},\
  }\href {\doibase 10.1103/PhysRevLett.101.026403} {\bibfield  {journal}
  {\bibinfo  {journal} {Phys. Rev. Lett.}\ }\textbf {\bibinfo {volume} {101}},\
  \bibinfo {pages} {026403} (\bibinfo {year} {2008})}\BibitemShut {NoStop}%
\bibitem [{\citenamefont {Yildirim}\ \emph {et~al.}(2001)\citenamefont
  {Yildirim}, \citenamefont {G\"ulseren}, \citenamefont {Lynn}, \citenamefont
  {Brown}, \citenamefont {Udovic}, \citenamefont {Huang}, \citenamefont
  {Rogado}, \citenamefont {Regan}, \citenamefont {Hayward}, \citenamefont
  {Slusky}, \citenamefont {He}, \citenamefont {Haas}, \citenamefont {Khalifah},
  \citenamefont {Inumaru},\ and\ \citenamefont {Cava}}]{Yildirim2001MgB2}%
  \BibitemOpen
  \bibfield  {author} {\bibinfo {author} {\bibfnamefont {T.}~\bibnamefont
  {Yildirim}}, \bibinfo {author} {\bibfnamefont {O.}~\bibnamefont
  {G\"ulseren}}, \bibinfo {author} {\bibfnamefont {J.~W.}\ \bibnamefont
  {Lynn}}, \bibinfo {author} {\bibfnamefont {C.~M.}\ \bibnamefont {Brown}},
  \bibinfo {author} {\bibfnamefont {T.~J.}\ \bibnamefont {Udovic}}, \bibinfo
  {author} {\bibfnamefont {Q.}~\bibnamefont {Huang}}, \bibinfo {author}
  {\bibfnamefont {N.}~\bibnamefont {Rogado}}, \bibinfo {author} {\bibfnamefont
  {K.~A.}\ \bibnamefont {Regan}}, \bibinfo {author} {\bibfnamefont {M.~A.}\
  \bibnamefont {Hayward}}, \bibinfo {author} {\bibfnamefont {J.~S.}\
  \bibnamefont {Slusky}}, \bibinfo {author} {\bibfnamefont {T.}~\bibnamefont
  {He}}, \bibinfo {author} {\bibfnamefont {M.~K.}\ \bibnamefont {Haas}},
  \bibinfo {author} {\bibfnamefont {P.}~\bibnamefont {Khalifah}}, \bibinfo
  {author} {\bibfnamefont {K.}~\bibnamefont {Inumaru}}, \ and\ \bibinfo
  {author} {\bibfnamefont {R.~J.}\ \bibnamefont {Cava}},\ }\href {\doibase
  10.1103/PhysRevLett.87.037001} {\bibfield  {journal} {\bibinfo  {journal}
  {Phys. Rev. Lett.}\ }\textbf {\bibinfo {volume} {87}},\ \bibinfo {pages}
  {037001} (\bibinfo {year} {2001})}\BibitemShut {NoStop}%
\bibitem [{\citenamefont {Wang}\ \emph
  {et~al.}(2017{\natexlab{b}})\citenamefont {Wang}, \citenamefont {Rademaker},
  \citenamefont {Dagotto},\ and\ \citenamefont
  {Johnston}}]{Wangyanprb2017linewidth}%
  \BibitemOpen
  \bibfield  {author} {\bibinfo {author} {\bibfnamefont {Y.}~\bibnamefont
  {Wang}}, \bibinfo {author} {\bibfnamefont {L.}~\bibnamefont {Rademaker}},
  \bibinfo {author} {\bibfnamefont {E.}~\bibnamefont {Dagotto}}, \ and\
  \bibinfo {author} {\bibfnamefont {S.}~\bibnamefont {Johnston}},\ }\href
  {\doibase 10.1103/PhysRevB.96.054515} {\bibfield  {journal} {\bibinfo
  {journal} {Phys. Rev. B}\ }\textbf {\bibinfo {volume} {96}},\ \bibinfo
  {pages} {054515} (\bibinfo {year} {2017}{\natexlab{b}})}\BibitemShut
  {NoStop}%
\bibitem [{\citenamefont {Zhou}\ and\ \citenamefont
  {Millis}(2017)}]{Millis2017prb}%
  \BibitemOpen
  \bibfield  {author} {\bibinfo {author} {\bibfnamefont {Y.}~\bibnamefont
  {Zhou}}\ and\ \bibinfo {author} {\bibfnamefont {A.~J.}\ \bibnamefont
  {Millis}},\ }\href {\doibase 10.1103/PhysRevB.96.054516} {\bibfield
  {journal} {\bibinfo  {journal} {Phys. Rev. B}\ }\textbf {\bibinfo {volume}
  {96}},\ \bibinfo {pages} {054516} (\bibinfo {year} {2017})}\BibitemShut
  {NoStop}%
\bibitem [{\citenamefont {Fuchs}\ and\ \citenamefont {Kliewer}(1965)}]{FK1965}%
  \BibitemOpen
  \bibfield  {author} {\bibinfo {author} {\bibfnamefont {R.}~\bibnamefont
  {Fuchs}}\ and\ \bibinfo {author} {\bibfnamefont {K.~L.}\ \bibnamefont
  {Kliewer}},\ }\href {\doibase 10.1103/PhysRev.140.A2076} {\bibfield
  {journal} {\bibinfo  {journal} {Phys. Rev.}\ }\textbf {\bibinfo {volume}
  {140}},\ \bibinfo {pages} {A2076} (\bibinfo {year} {1965})}\BibitemShut
  {NoStop}%
\end{thebibliography}

\end{document}